\documentclass[twocolumn,showpacs, prd, nofootinbib,superscriptaddress,amsmath,amssymb,floatfix]{revtex4-1}
\usepackage{graphicx,color}  % needed for figures
\usepackage{dcolumn}   % needed for some tables
\usepackage{bm}        % for math
\usepackage{amssymb}   % for math
\usepackage{lineno, soul}
\usepackage[dvipsnames]{xcolor}
\usepackage{color}
\usepackage{natbib}
\usepackage[normalem]{ulem}
\usepackage{booktabs,multirow}
\usepackage{caption}
\usepackage{subcaption}
\usepackage{placeins}
\usepackage{enumitem}

\def \aap  {A\&A}

\def \apjl  {ApJL}

\def \apss  {AP\&SS}

\begin{document}

% The following information is for internal review, please remove them for submission
\widetext
%\leftline{Version 00 as of \today}
%\leftline{Comment to {\tt d0-run2eb-nnn@fnal.gov} by xxx, yyy}

% the following line is for submission, including submission to the arXiv!!
%\hspace{5.2in} \mbox{Fermilab-Pub-04/xxx-E}

\title{New estimate for the contribution of the Geminga pulsar to the positron excess} 

\author{Donglin Wu}
\affiliation{Shanghai World Foreign Language Academy}
\email{donglinwu.michael@gmail.com}

\date{\today}

\begin{abstract}
The origin of the positron excess is one of the most intriguing mysteries in astroparticle physics.
The recent discovery of extended $\gamma$-ray halos around the pulsars Geminga, Monogem and PSR J0621+3755 have brought indirect evidence that pulsar wind nebulae accelerate $e^{\pm}$ up to very-high-energy.
While the precision of previous data does not permit precise evaluation of the parameters for the pulsars, we are able to find the more precise shape of the injection spectrum using new data released by HAWC and LHAASO in 2020 and 2021.
%we investigate the origin of the extended $\gamma$-ray emission around pulsars to find the shape of the injection spectrum more precisely. 
We find that this is well fitten by a power-law with an exponential cutoff. The spectral index is quite hard with values around 1 while the cutoff energy is roughly 100 TeV.
We also derive the strength of the diffusion coefficient around the pulsars finding that it is two orders of magnitude lower than the average of the Galaxy.
Finally, we use the above mentioned results to estimate the contribution of Geminga to the positron excess. This source alone can contribute to the entire positron excess at around 1 TeV.

\end{abstract}

\maketitle

\section{Introduction}

%\textcolor{blue}{Change all positrons into $e^+$}

An excess of positrons ($e^+$) above 10 GeV has been detected for the first time by Pamela~\cite{Pamela}, Fermi-LAT~\cite{Fermi-LAT} and then confirmed with unprecedented precision by AMS-02~\cite{AMS1, AMS2}.
This excess, called positron excess, is present with respect to the well known secondary production (for example, refer to Ref. \cite{Moskalenko_1998}) that is due to the collision of cosmic rays against the atoms of the interstellar medium, primarily consisted of hydrogen and helium atoms.
The origin of the positron excess is still a mystery. Different processes and sources have been invoked as possible solutions to this puzzle.

One possible astrophysical interpretation is related to supernova remnants (SNRs) from which $e^+$ can be produced through the collision of cosmic rays accelerated in the SNR shock wave with atoms still present in the remnant~\cite{Blasi_2009,Ahlers_2009, Mertsch_2014, Cholis_2014, Tomassetti_2015, Mertsch_2021}.
An exotic process related to new particles, called dark matter, has been also theorized as a possible solution to this puzzle~\cite{Cholis_2009, Cholis_2013}. Dark matter particles are weakly interacting massive particles that make up $80\%$ of the total matter in the Universe and they could be made of new particle with respect to the ones of the Standard Model \cite{Bertone_2005}.
However, the dark matter candidates that explain the positron excess would have a too large annihilation cross sections to be compatible with the no detection of any signal from other cosmic particles. Indeed, if dark matter explain the positron excess a bright signal of other messengers for example $\gamma$ rays should be produced too and be detected for example from the center of the Galaxy or dwarf Spheroidal Galaxies of the Milky Way where the density of dark matter is very large \cite{Meade_2010}. The fact that probes of Fermi-LAT and MAGIC did not detect significant $\gamma$-ray halos in most of these regions challenges this interpretation \cite{Abdo_2010, Ackermann_2012, Acciari_2022, Gammaldi_2021}, although the $\gamma$-ray emissions from the Galactic center can be consistent with it \cite{Daylan_2016}.

Finally, Galactic pulsars, rapidly rotating neutron stars, have been claimed to explain the positron excess using a few up to tens percent of their total energetics~\cite{Linden_2013, Di_Mauro_2014, Cholis_2013, Ibarra_2010, Hooper_2009, Pato_2010, Boudaud_2015, Mauro_2016, Hooper_2017}.
A pulsar loses part of its rotational energy in the form of pulsar winds created by its strong magnetic field. Pulsar Wind nebulae (PWNe) are nebulae of relativistic $e^{\pm}$ powered by pulsar winds and balanced in pressure by ambient environment of supernova remnants. Through interaction with SNR medium, PWNe is thought to be able to accelerate $e^+$ up to very-high-energy and inject them into interstellar medium \cite{Chi_1996, Gaensler_2006, Kargaltsev_2017, Reynolds_2017, Amato_2014, Blasi_2010}.

Various measurements may provide indirect evidence that $e^+$ accelerated and injected by pulsars are responsible for the positron excess. The High-Altitude Water Cherenkov (HAWC) observatory, one of the world's most sensitive wide-field multi-TeV observatories, has detected extended $\gamma$-ray emission, of the order of a few degrees, between 8-40 TeV around two nearby pulsars, Geminga and PSR B0656+14, as known as Monogem \cite{HAWC2017} (labelled hereafter HAWC2017). 
A very extended $\gamma$-ray halo with significance $7.8-11.8\sigma$ has been also found around Geminga at energies above 8 GeV in the data measured by the Fermi Large Area Telescope (Fermi-LAT) \cite{Mattia_2019}. 

These photons could be produced by $e^{\pm}$ accelerated by the PWN and inverse Compton scattering (ICS) against the interstellar radiation field (ISRF) low-energy photons. Studies have shown that the model for such scattering is compatible with the $\gamma$-ray data if the diffusion coefficient around the PWNe is between two and three orders of magnitude lower than the Galactic average, which is derived through the ratio of primary to secondary cosmic rays \cite{Vecchi_2022, Weinrich_2020, Korsmeier_2021}. A possible interpretation for these $\gamma$-ray halos is that PWNe are surrounded by small regions with inhibited diffusion coefficient while the rest of the Galaxy maintains an average diffusion of the order of $10^{28}~\mathrm{cm}^2~\mathrm{s}^{-1}$ \cite{Hooper_2017, Profumo_2018, Evoli_2018, Tang_2019, Mattia_2019, Gulaugur_2019}. Several authors have analyzed and explored this two-zone diffusion model for $e^+$ showing that it can explain well the morphology of the $\gamma$-ray emission around Geminga and Monogem and show the consequences for the positron excess \cite{Evoli_2018, Fang_2018, Tang_2019, Gulaugur_2019, Mattia_2019}. Ref.~\cite{Evoli_2018, Recchia_2021, Mukhopadhyay_2021} discussed possible theoretical explanation for inhibition of diffusion near pulsars. In particular Refs.~\cite{Evoli_2018, Mukhopadhyay_2021} propose that streaming instabilities generated by the cosmic rays accelerated by the PWNe could reduce significantly the diffusion around these sources. Instead ref.~\cite{Recchia_2021} show that the inclusion of the ballistic propagation that occurs before the diffusive one could explain the morphology of the $\gamma$-ray halos with a diffusion coefficient strength of the same order of the one needed to fit cosmic-ray data.

Recent studies have also evaluated diffusion around pulsars in greater details. Ref.~\cite{Martin_2022} has modeled the diffusion halos around pulsars Geminga and B0656+14 in different scenarios using measurements from Fermi-LAT and HAWC, and found that most middle-aged pulsars may not develop halos using the AMS-02 data and population synthesis. Further works in Ref. \cite{Martin_2022_2} suggest that if diffusive halos are rare, either better PWNe modeling is required or another type of unknown sources is responsible for very-high-energy $\gamma$-ray emission. Ref. \cite{Luque_2022} has analyzed that the alternate diffusive model, anisotropic diffusion model, cannot explain the size and radial symmetry of $\gamma$-ray halos while being compatible with values derived from Galactic primary-to-secondary ratios.

Although injection from PWNe is a promising explanation for $e^+$ excess, the data provided by HAWC do not permit to estimate the size of the low-diffusion bubble. In fact, at TeV energies the size of $\gamma$-ray halos is mainly driven by the energy losses instead of diffusion, i.e. the former have much smaller time scale of the latter at these very-high-energies \cite{Evoli_2017}.
Moreover, the positron spectral index is not precisely determined because the data from the detection in the Fermi and HAWC bands are not precise enough. For example Ref. \cite{Mattia_2019} find that a combined fit to the Fermi-LAT and HAWC gives between 1.8 and 2.2. 
In addition, the data from HAWC released in Ref. \cite{Escobedo_2021} are a factor of 3, larger than HAWC2017 so the spectral index must be redetermined.
These two facts bring a large source of uncertainty in the prediction of the positron flux and that the contribution of pulsars such as Geminga to the positron excess deserves more investigation.

Fortunately, new data has been released since 2017. In 2020, HAWC released new data for the surface brightness of Geminga in three energy bins, above 1 TeV, 5.6-17 TeV and 17-56 TeV \cite{Zhou_2020}. Proceedings in 2021 provided the energy spectrum of $\gamma$-rays near Geminga measured  by HAWC in the range of 0.8-66 TeV \cite{Escobedo_2021}. These new data may permit more precise estimation of the $e^+$ flux from Geminga. Recently, the Large High Altitude Air Shower Observatory (LHAASO), the most sensitive observatory with gamma-rays above 10 TeV, measured the surface brightness and energy spectrum of $\gamma$-rays surrounding pulsar J0621+3755 \cite{Aharonian_2021}. The investigation of pulsar J0621+3755 helps determine whether the assumptions adopted for Geminga can apply to other pulsars in general.

In this paper, we use the new data, to estimate the contribution of Geminga to positron excess, after obtaining the best-fit physical parameters. With the new data of HAWC and LHAASO combined with measurement of Fermi-LAT, we are now able to constrain the spectral index $\gamma_e$ as well as the cutoff energy $E_c$ of the $e^+$ injection spectrum for Geminga and J0621+3755 for different continuous injection scenario with different characteristic spin-down timescale $\tau_0$. We find that the best-fit range of $E_c$ and $\tau_0$ of the pulsars are consistent, being 90-120 TeV and 10-30 kyr respectively, while $\gamma_e$ is around $0.7-1.3$, about $50\%$ smaller than those in previous studies \cite{Mattia_2019}. We conduct fit to the surface brightness of Geminga and J0621+3755, finding diffusion coefficient compatible with previous results. Although disagreement of $D_0$ for different energy bins for Geminga is found, no firm conclusion can be made considering the large errors. At last, we estimate the $e^+$ flux contributed by Geminga using best-fit parameters. We predict that $e^+$ flux with energy around 1 TeV would have a sharp spike due to the contribution of Geminga. The notable novelty of this work is that for the $\gamma$-ray energy spectrum and surface brightness of pulsars, we use new data released by HAWC for Geminga and new data from LHAASO for J0621+3755, which allows us to constrain the diffusion coefficients, pulsar parameters and injection spectrum of $e^{\pm}$. This in turn allows us to compute the new estimate of the contribution of Geminga to $e^+$ excess.

The rest of the paper is organized as follows. Section \ref{Model} describes the model used for injection of $e^{\pm}$ from PWNe, the propagation of $e^{\pm}$ in Galaxy and the gamma ray production from ICS between the injected $e^{\pm}$ and ISRF. Section \ref{sec:result} presents the main results and discusses their physical interpretations. Section \ref{sec:conclu} summarizes the paper.

\section{Model}
\label{Model}

\subsection{Positron and electron injection from pulsar wind nebulae}\label{Source}

Highly magnetized pulsars are known source of $e^{\pm}$ in the Milky Way. The rotational energy of pulsars is carried away in the form of a magnetized wind generated by strong magnetic field produced by the pulsar. The PWNe comprised of relativistic $e^{\pm}$ pair is powered by the pulsar wind, while being confined by local SNR environment. When the SNR shock wave interacts with the interstellar medium, a reverse shock is generated. When this reverse shock hits the PWNe, the $e^{\pm}$ from the PWNe are able to exit the nebulae and be injected into the interstellar medium \cite{Amato_2014, Chi_1996, Blasi_2010}.

Assuming that all the particles are injected at the age of the source ($T$), a commonly-used burst-like injection spectrum $Q(E)$ is given by a power-law with an exponetial cutoff \cite{Delahaye_2010, Ellison_2007, Malkov_2001}:
\begin{equation}
    Q(E) = Q_0 \left(\frac{E}{E_0}\right)^{-\gamma_{e}} \exp\left(-\frac{E}{E_c}\right),
\end{equation}
where $Q_0$ is the normalization constant in unit of GeV$^{-1}$, $ \gamma_{e} $ is the slope of the spectrum and $E_c$ is the cutoff energy. The cutoff energy of the $e^{\pm}$ injection spectrum should be larger than 50 TeV to produce $\gamma$ rays with tens of TeV as observed by HAWC around Geminga and Monogem. The value of $\gamma_e$ found in previous studies are usually around 2.0 \cite{Ellison_2007, Mattia_2019}. \par

Indeed, the injection spectrum of $e^+$ is better modeled by a broken power law with a break around a few hundred GeV, an index harder than 2 below and an index softer than 2 above \cite{Bucciantini_2010, Torres_2014, Hess_2019, Principe_2020, Evoli_2021}. However, the $\gamma$-ray emissions we will analyze poorly constrain the index below the break, as the $\gamma$-ray energies are above 10 GeV, which require mostly $e^+$ with energies beyond 1 TeV \cite{Di_Mauro_2021}. Thus, we adopt a single power-law with cutoff instead of a broken power-law. However, it should be noted that using a single power law will affect the estimation of total energy of $e^+$ injected.

The total energy of $e^{\pm}$ for the burst-like injection spectrum in units of GeV is calculated as:
\begin{equation}
    E_{\rm{tot}} = \int_{E_{\rm{min}}}^{\infty}\,dE~E~Q(E) ,
\end{equation}
where $E_{\rm{min}}$ is the typical value of minimum of energy of electrons, 0.1 GeV. The normalization constant of the spectrum $Q_0$ can be derived assuming that a fraction $\eta$ of the total energy is converted into $e^{\pm}$:
\begin{equation}
    \label{E_tot norm}
    E_{\rm{tot}} = \eta W_0,
\end{equation} 
where $\eta$ is the efficiency of the conversion of spin-down energy of pulsars $W_0$ to energy of $e^{\pm}$ injected. $W_0$ represents the energy released by the pulsar from its birth to the age of the pulsar $T$ and is calculated as:
\begin{equation}
    \label{W0}
    W_0 = \tau_0 \dot{E}~\left(1 + \frac{T}{\tau_0}\right)^2.
\end{equation}
$\tau_0$ is the characteristic pulsar spin-down timescale.
As we will show in the next section values of $\tau_0$ larger than at least a few kyrs are needed.
The spin-down luminosity $\dot{E}$, the rate at which rotational energy is lost from the pulsar, the observed age $t_{\rm{obs}}$ as well as the distance $d$ of the pulsars are obtained from the ATNF catalog \cite{ATNF}, if not stated otherwise. The actual age $T$ used in equation \ref{W0} is calculated through the equation $T = t_{\rm{obs}} + d/c$.  \par
For Geminga, we use $d=250$ pc, $t_{obs}=370$ kyr and $\dot{E} = 3.2 \times 10^{34}$ erg. For pulsar J0621+3755, we use $d=1.6$ kpc, $t_{obs}=207.8$ kyr and $\dot{E} = 2.7 \times 10^{34}$ erg \cite{Aharonian_2021}. Since the pulsars are relatively close to Earth, the actual age and observed age do not differ significantly. However, for farther source, the distinction could be relevant for the model.  \par

Notice that $\eta$ in equation \ref{E_tot norm} is the efficiency of conversion of energy into $e^{\pm}$. When we calculate the flux on Earth of $e^+$, the actual efficiency used in the calculation is $\eta_{+} = \eta/2$, as we assume an equal number of positrons and electrons injected from pulsars

The burst-like injection scenario is not able to predict the existence of multi TeV photons. The reason behind this is related to energy losses suffered by $e^{\pm}$ after being injected from the source after its creation. For example $e^{\pm}$ emitted 370 kyrs ago from Geminga could reach the Earth with a maximum energy of about $E_{\rm{max}} \approx 1/(b(E)T) = 1.3$ TeV where $b(E)$ are the energy losses.
Compared to the burst-like scenario, the continuous injection scenario includes a continuous injection of $e^{\pm}$ in time. The time dependence is intrinsic in the variation of the spin down luminosity across the pulsar age. The continuous injection spectrum of $e^{\pm}$ is given by:
\begin{equation}
    Q(E, t) = L(t) \left(\frac{E}{E_0}\right)^{-\gamma_{e}} \exp\left(-\frac{E}{E_c}\right),
\end{equation}
where $L(t)$ is the magnetic dipole braking:
\begin{equation}
    L(t) = \frac{L_0}{\left(1 + \frac{t}{\tau_0}\right)^2}.
\end{equation}
The total energy of $e^{\pm}$ injected from a pulsar with age $T$ is found by integrating the injection spectrum in energy and time: 
\begin{equation}
    E_{\rm{tot}} = \int_{0}^{T} \,dt \int_{E_{\rm{min}}}^{\infty} \,dE~ E~Q(E,t),
\end{equation}
and the normalization value $L_0$ can be obtained by equation \ref{E_tot norm}, similarly to the burst like model.\par

\subsection{Propagation of electrons and positrons}
\label{sec:propagation}
$e^{\pm}$ propagate in the Galaxy after being injected by the PWNe. The propagation is a very complex process as $e^{\pm}$ are subject to energy losses and diffusion on the Galactic magnetic field irregularities and secondary production due to collision of cosmic rays against interstellar medium (ISM) atoms. For high-energy $e^{\pm}$, the usual propagation conservation equation can be written as \cite{Delahaye_2010}:
\begin{equation}
\label{eq:prop}
    \partial_{t}\mathcal{N}-\nabla\cdot\{D(E)\nabla\mathcal{N}\}+\partial_{E}\left(b(E)\mathcal{N}\right)~=~\mathcal{Q}(E, \mathbf{x}, t),
\end{equation}
where $\mathcal{N}\equiv\mathcal{N}(E, \mathbf{x}, t)=dn/dE$ is the electron number density per unit of energy. $D(E)$ is the energy-dependent diffusion written, for a single diffusive zone with infinite radius, as:
\begin{equation}
\label{eq:diff}
    D(E)=D_0(E/\mathrm{1 GeV})^{\delta},
\end{equation}
where $D_0$ is the diffusion coefficient at 1 GeV and $\delta$ is the slope. We adopt $\delta = 1/3$ throughout the paper otherwise differently stated. 
$\mathcal{Q}(E, \mathbf{x}, t)$ is the source term, i.e. the number of $e^{\pm}$ injected from the PWNe, reported in section \ref{Source}. $b(E)$ represents the energy losses of $e^{\pm}$, which adopts the shape of a power law on first approximation, i.e. $b(E)\propto E^{\alpha}$. The energy losses consist mainly of two processes. The first one is due to $e^{\pm}$ inverse Compton scattering on the low-energy photons of the ISRF, whose slope with respect to energy is roughly $\alpha=1.9$ \cite{ISRF_2021}. The second process is the Synchrotron radiation that positrons and electrons encounter while propagation in the Galactic magnetic field. The slope of the Synchrotron losses is equal to 2 and constant with energy. Unless otherwise stated, we adopt the spectrum of the local ISRF described in Ref. \cite{Vernetto:2016alq} and a Galactic magnetic field of strength 3 $\mu$G for the energy losses. 
%To keep the form of power law to derive analytical solutions from conservation equation, we use a four-section power law to approximate the energy loss described in Vernetto 2016. \par

Using the propagation equation, in a homogeneous diffusive halo with infinite radius, the flux for $e^{\pm}$ of specific energy $E$ at position $\mathbf{r}$ under a burst-like scenario is given by \cite{Delahaye_2009}:
\begin{equation}
    \mathcal{N}(E, \mathbf{r}) = \frac{b(E_s)}{b(E)} \frac{1}{(\pi \lambda^2)^{\frac{3}{2}}}~\exp\left(-\frac{\mid\mathbf{r}-\mathbf{r_s}\mid}{\lambda^2}\right)Q(E_s),
\end{equation}
where $\mathbf{r_s}$ is the position of the source, and $\lambda$ is the propagation scale length that describes the typical distance traveled by a $e^{\pm}$ under the influence of energy losses and diffusion. 
Considering an electron emitted at the PWNe with an energy $E_s$ and detected on Earth at energy $E$ the propagation length is expresses as:
\begin{equation}
    \lambda^2=\lambda^2(E, E_s)=4\int^{E_s}_{E} \frac{D(E')}{b(E')} \,dE'.
\end{equation} 
%during a period of time $\Delta \tau$: 
%\begin{equation}
%    \Delta \tau \equiv \int^{E_s}_{E} \frac{dE'}{b(E')} = t - t_{obs}.
%\end{equation}

In a model where the diffusion and the energy losses are homogeneous in the Galaxy, the flux for $e^{\pm}$ with energy $E$, at position $\mathbf{r}$ and time $t$ under a continuous injection is given by \cite{Yuksel_2009}:
\begin{multline}
    \mathcal{N}(E, \mathbf{r}, t) = \int^{t}_{0} \frac{b(E_s(t'))}{b(E)} \frac{1}{(\pi \lambda(t', t, E)^2)^{\frac{3}{2}}} \times \\
    \times ~\exp\left(-\frac{\mid\mathbf{r}-\mathbf{r_s}\mid}{\lambda(t', t, E)^2}\right)Q(E_s(t')) \, dt',
\label{eq:onezoneN}
\end{multline}
where the integration over time accounts for the continuous injection of $e^{\pm}$ from PWNe. \par

Data from HAWC 2017 suggests that a two-zone diffusive model may be in demand to derive the flux of $e^+$ observed on Earth: the pulsars are surrounded by a diffusive halo with low diffusion coefficient, which is surrounded by regions with Galactic average diffusion coefficient. A commonly used model is to implement the two-zone energy-dependent diffusion factor:
\begin{equation}
    D(E, r)=
    \begin{cases}
    D_0(E/\mathrm{1 GeV})^{\delta}, & \text{for}\ 0<r<r_b,\\
    D_2(E/\mathrm{1 GeV})^{\delta}, & \text{for}\ r\geq r_b,
    \end{cases}
\end{equation}
where $r_b$ is the radius of inner low diffusion zone. The flux of $e^{\pm}$ for the continuous injection scenario in equation \ref{eq:onezoneN} becomes \cite{Tang_2019}:
\begin{equation}
    \mathcal{N}(E, \mathbf{r}, t) = \int^{t}_{0} \, dt' \frac{b(E(t'))}{b(E)}Q(E(t'))\mathcal{H}(\mathbf{r}, E),
\label{eq:twozoneN}
\end{equation}
where the term $\mathcal{H}(\mathbf{r}, E)$ is given by:
\begin{widetext}
\begin{multline}
    \label{Two-zone-flux}
    \mathcal{H}(\mathbf{r}, E) = \frac{\xi(\xi+1)}{\pi^{3/2}\lambda_0^3[ 2\xi^2\mathrm{erf}(r_b/\lambda_0)-\xi(\xi-1)\mathrm{erf}(2r_b/\lambda_0)+2\mathrm{erfc}(r_b/\lambda_0)]} \times \\
    \times
    \begin{cases}
    e^{\left( -\frac{\Delta r^2}{\lambda_0^2} \right)} + \left(\frac{\xi-1}{\xi+1}\right)\left( \frac{2r_b}{\Delta r} \right)e^{\left( -\frac{(\Delta r-2r_b)^2}{\lambda_0^2} \right)}, & \ 0<r<r_b,\\
    \left(\frac{2\xi}{\xi+1}\right) \bigg[ \frac{r_b}{\Delta r} + \xi \left(1-\frac{r_b}{\Delta r}\right) \bigg]e^{\left( -\big[\frac{(\Delta r-r_b)}{\lambda_2}+\frac{r_b}{\lambda_0}\big]^2 \right)}, & \ r\geq r_b,
    \end{cases}
\end{multline}
\end{widetext}
where $\Delta r = |\mathbf{r}-\mathbf{r_s}|$ is the distance to the source, b is defined as $\xi = \sqrt{D_0/D_2}$, $\lambda_0$ and $\lambda_2$ are typical propagation scale length in region with diffusion coefficient $D_0$ and $D_2$. Mathematically, when $D_0$ is equal to $D_2$ or when $r_b$ approaches infinity, equation \ref{Two-zone-flux} reduces to equation \ref{eq:onezoneN}. \par 

\subsection{$\gamma$-ray emission for inverse Compton scattering}

The high-energy $\gamma$ rays observed in the direction of the pulsars are generated when very energetic $e^{\pm}$ injected by the PWNe scatter with low-energy photons from ISRF through ICS in the Klein–Nishina regime \cite{Lyutikov_2012}. 

The photon flux from ICS at energy of $E_{\gamma}$ and for a solid angle $\Delta \Omega$ is given by \cite{ICS_1970, ICS_2011}:
\begin{multline}
    \label{ICS_flux}
    \phi^{IC}(E_{\gamma}, \Delta \Omega) = \frac{1}{4\pi} \int_{m_{e} c^2}^{\infty} \,dE ~\mathcal{P}^{IC}(E, E_{\gamma}) \times \\
    \times \bigg[ \int_{\Delta \Omega} \,d\Omega \int_{0}^{\infty} \,ds~ \mathcal{N}_e(E, \mathbf{r}(s,\Omega), T) \bigg] ,
\end{multline}
where $\mathcal{N}_e(E, \mathbf{r}(s,\Omega), T)$ is the energy spectrum of $e^{\pm}$ with energy $E$ at a distance $r$ in the Galaxy. $\mathcal{N}_e(E, \mathbf{r}(s,\Omega), T)$ comes from the solution of the propagation equation assuming a one Eq.~\ref{eq:onezoneN} or two zone model (see Eq.~\ref{eq:twozoneN}).
%For a pulsar with age $T$, this parameter is given by:
%\begin{equation}
%\label{eq:specpos}
%    \mathcal{N}_e(E, \mathbf{r}(s, \Omega)) = \int_{0}^{T} \mathcal{N}_e(E, \mathbf{r}(s, \Omega), t) \, dt,
%\end{equation}
%where $\mathcal{N}_e(E, \mathbf{r}(s, \Omega), t)$ is the time-dependent flux, the solution of propagation equation (equation \ref{eq:prop}), and $\mathbf{r}(s, \Omega)$ is the position as a function of line of sight $s$ and angle from the source $\Omega$. 
The latter term in square bracket in equation \ref{ICS_flux}, the integral of $\mathcal{N}_e(E,\mathbf{r}(s, \Omega))$ along the line of sight within a solid angle $\Delta \Omega$, represents the spectrum of $e^{\pm}$ propagating through the Galaxy from a solid angle $\Delta \Omega$. 
Since the emission from the source is assumed to be spherically symmetric the solid angle $\Delta \Omega$ can be parametrized as a function of the angle $\theta$ that represents the angular separation with respect to the line of sight that points to the center of the source.

$\mathcal{P}^{IC}(E, E_{\gamma})$ is the power of photons with energy $E_{\gamma}$ emitted by the scattering of a single $e^{\pm}$ with energy $E$, and is given by:
\begin{multline}
    \mathcal{P}^{IC}(E, E_{\gamma}) = \frac{3\sigma_T m_e^2 c^5}{4E^2} \int_{\frac{m_e c^2}{4E}}^{1} \,dq~ \frac{\mathcal{N}}{d\epsilon}(\epsilon(q)) \times \\
    \times \left( 1-\frac{m_e^2 c^4}{4qE(E-E_{\gamma})} \right) \bigg[ 2q \log q
    +q+1-2q^2+ \frac{E_{\gamma}(1-q)}{E-E_{\gamma}} \bigg],
\end{multline}
where $m_e$ is the electron mass, $\epsilon$ is the ISRF photon energy, $q$ is a variable defined for convenience of integration:
\begin{equation}
    q = \frac{E_{\gamma} m_e^2 c^4}{4\epsilon(E-E_{\gamma})},
\end{equation}
and $\frac{d\mathcal{N}}{d\epsilon} (\epsilon(q))$ is the energy spectrum of ISRF, as described in Refs. \cite{Vernetto:2016alq, ISRF_2021}.

The model has different parameters. The most important ones are related to the strength and the size of the low-diffusion bubble ($D_0$, $D_2$, $r_b$). These parameters could change the results of the positron flux on Earth and the observed $\gamma$-ray flux. Unfortunately, current data and available papers are not sensitive to the exact shape of the low-diffusion bubble and the exact transition between the two diffusive zones. Due to above reasons we decided to implement the two-zone model in equation \ref{Two-zone-flux} only for the calculation of the positron flux on Earth while for the to ICS model in equation \ref{ICS_flux} we use the one zone model. 

While the energy spectrum of $\gamma$-ray provides information about the injection spectrum of $e^{\pm}$, the variation of $\gamma$-ray flux with angle from the source provides information about the propagation of $e^{\pm}$. The surface brightness of a pulsar is defined as:
\begin{multline}
\label{eq:ICS_SB}
    \Phi^{IC}(E_{\rm{min}}, E_{\rm{max}}, \Omega) = \frac{c}{4\pi} \int_{E_{\rm{min}}}^{E_{\rm{max}}} \, dE_{\gamma} ~ \int_{m_{e} c^2}^{\infty} \,dE \\ \mathcal{P}^{IC}(E, E_{\gamma})
    \times \bigg[ \int_{0}^{\infty} \,ds~ \mathcal{N}_e(E, \mathbf{r}(s, \Omega), T) \bigg],
\end{multline}
where $E_{min}$ and $E_{max}$ give the range of energy of $\gamma$-ray measured. 

In this work, we do not consider the proper motion of the pulsars, because it only significantly affects the $\gamma$-ray emission morphology of pulsars at low energies. For example, the proper motion of Geminga, at $178.2\pm1.8$ mas/year \cite{Faherty_2007}, affects $\gamma$-ray emission with energy smaller than 100 GeV \cite{Mattia_2019}. On the other hand, the surface brightness of Geminga measured by HAWC in 2020 is for energies above 1 TeV, while that of pulsar J0621+3755 measured by LHAASO in 2021 is for energies above 25 TeV. These measurements at high energies are barely affected by the proper motions.

\FloatBarrier

\section{Results}
\label{sec:result}
\subsection{Variation of parameters for $\gamma$-ray energy spectrum}
\label{sec:varp}

%Section \ref{Model} describes the model of the emission of $\gamma$-ray through ICS of $e^{\pm}$ injected from pulsars and ISRF, including the definition of two important functions surface brightness and energy spectrum. 
In this section, we analyze how the physical parameters described in the model affect the energy spectrum of $\gamma$-ray. 
In particular we consider the parameters of $e^{\pm}$ injection spectrum, the spectral index $\gamma_e$ and the cutoff energy $E_c$, the energy losses, parameterized as $b_0 E^\alpha$, and $\tau_0$ the typical pulsar decay time.

    \begin{figure*}
        \centering
        \begin{subfigure}[b]{0.475\textwidth}
            \centering
            \includegraphics[width=\textwidth]{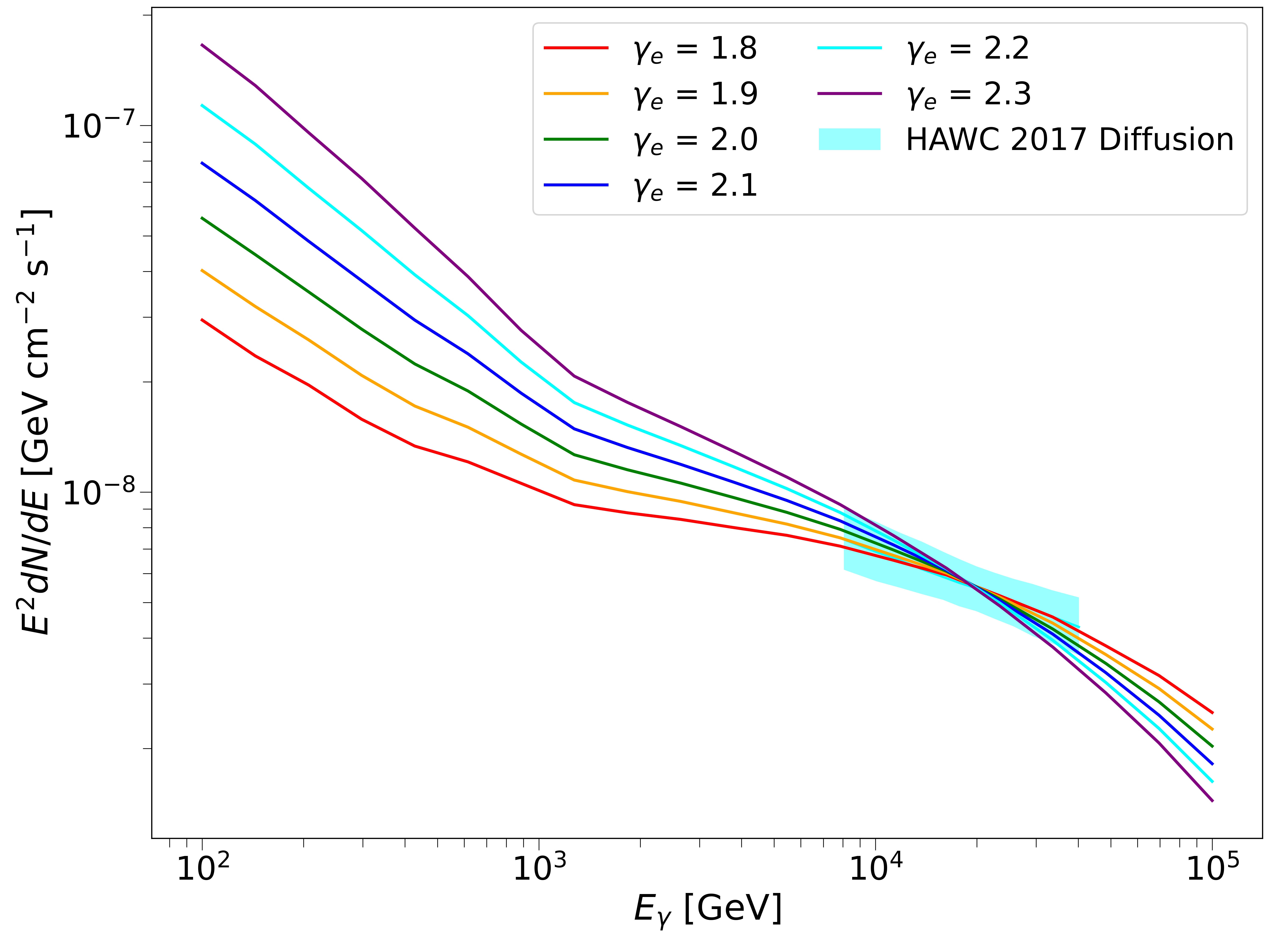}
        \end{subfigure}
        \hfill
        \begin{subfigure}[b]{0.475\textwidth}  
            \centering 
            \includegraphics[width=\textwidth]{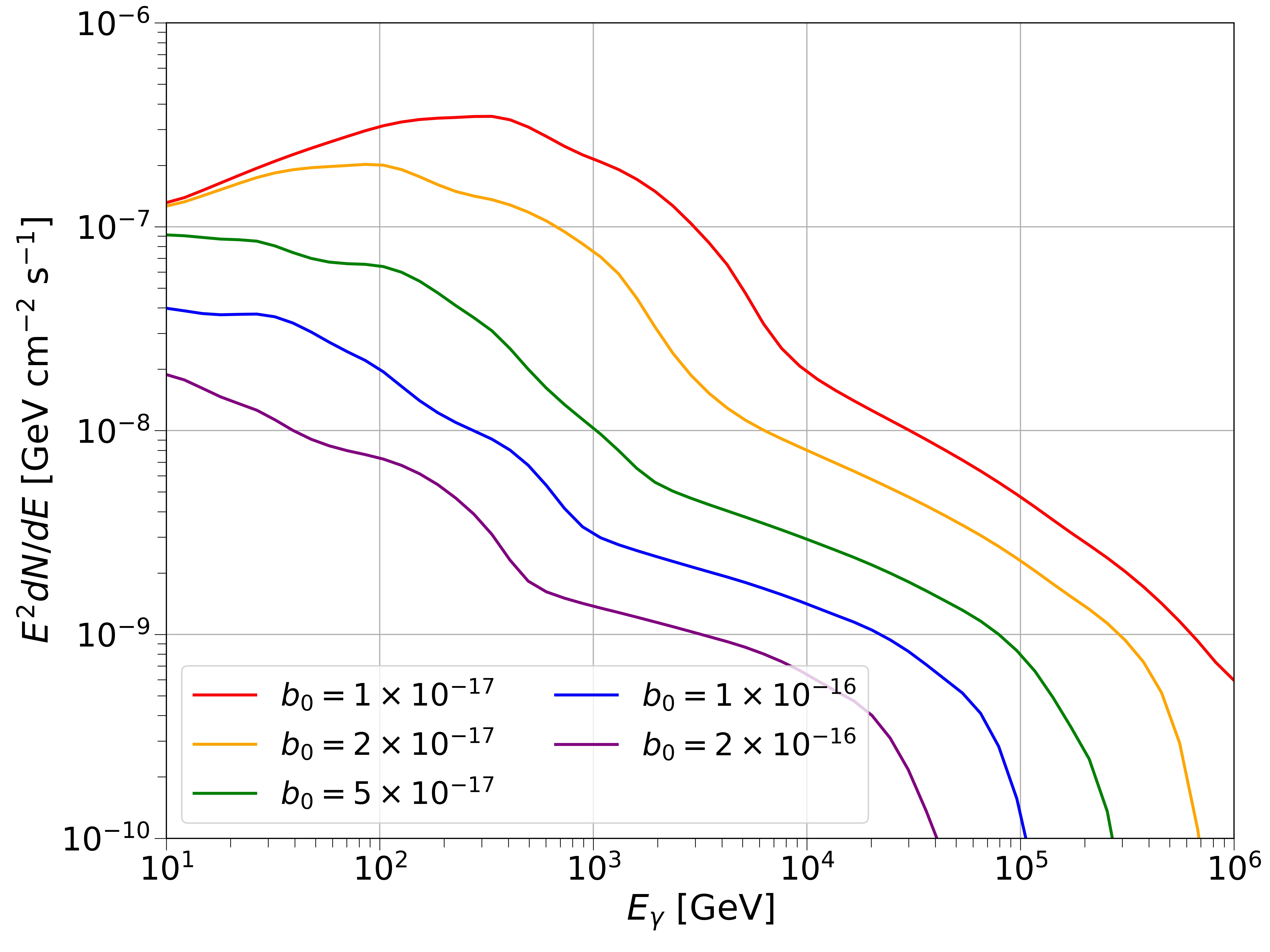}
        \end{subfigure}
        \vskip\baselineskip
        \begin{subfigure}[b]{0.475\textwidth}   
            \centering 
            \includegraphics[width=\textwidth]{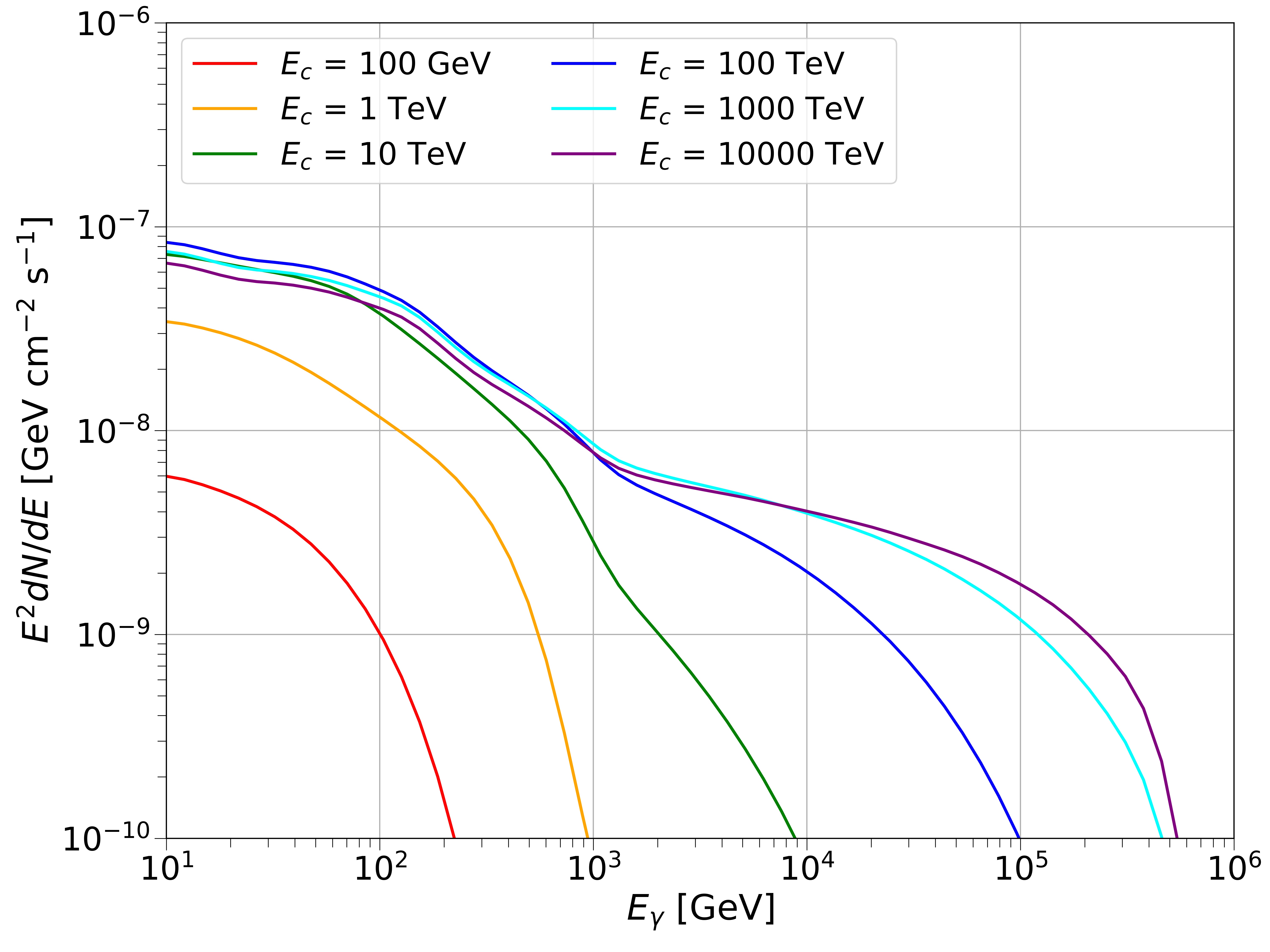}
        \end{subfigure}
        \hfill
        \begin{subfigure}[b]{0.475\textwidth}   
            \centering 
            \includegraphics[width=\textwidth]{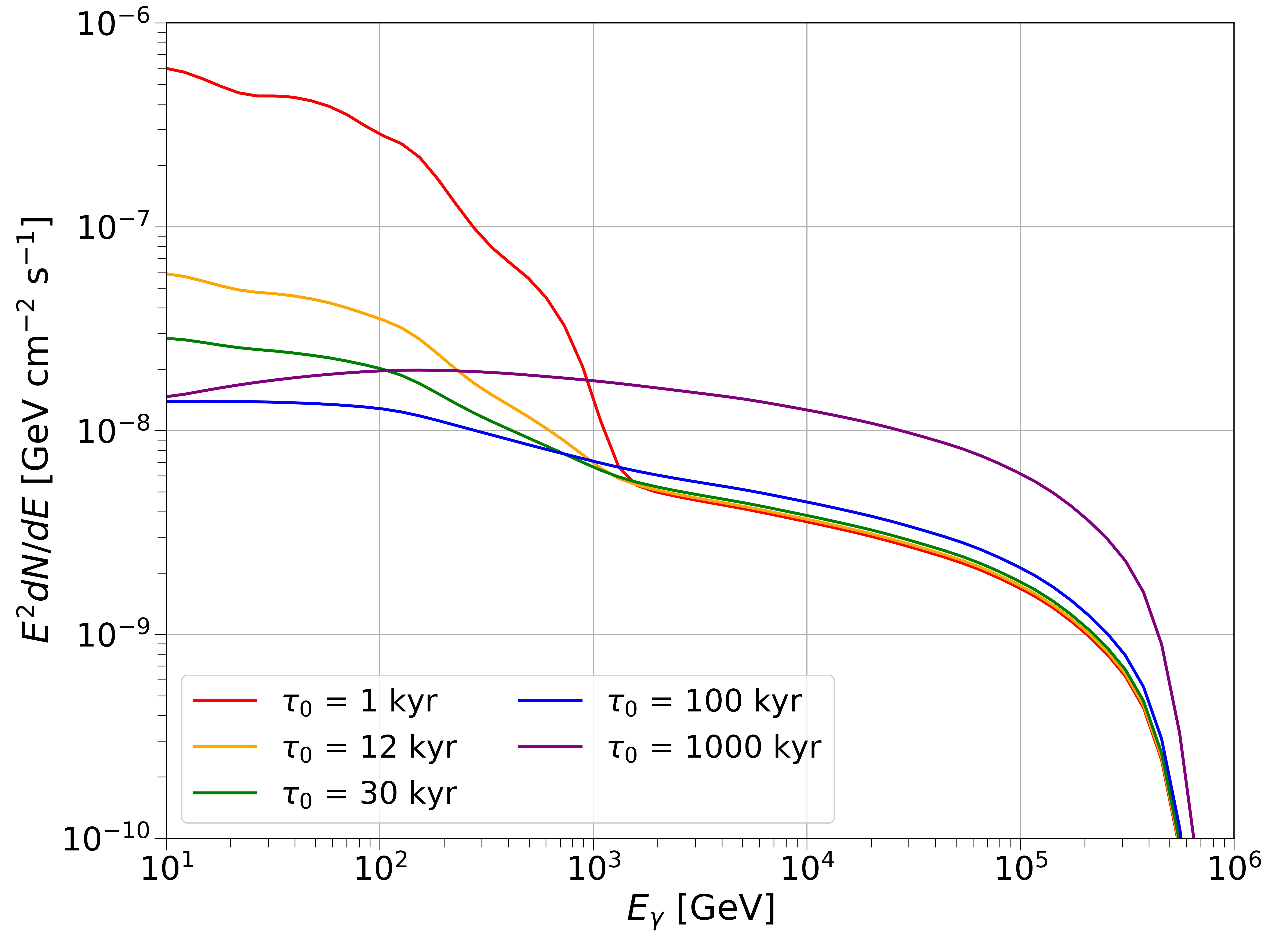}
        \end{subfigure}
        \caption[]
        {\small $\gamma$-ray flux of Geminga found by varying the physical parameters $\gamma_e$, $b_0$, $E_c$, $\tau_0$. The benchmark model adopts $\delta=1/3$, $D_0 = 1.0 \times 10^{26} \mathrm{cm}^2~\mathrm{s}^{-1}$, $\theta = 10$ degrees, $\tau_0 = 12$ kyr, $E_c = 1000$ TeV, $\gamma_e = 2.0$, $\eta=0.10$, and energy losses described in Ref.~\cite{Vernetto:2016alq}. Top left: Best-fit $\gamma$-ray flux to HAWC2017 data for spectral indexes $\gamma_e$ from 1.8 to 2.3. The efficiency values obtained for each $\gamma_e$ are 0.084, 0.118, 0.198, 0.406, 0.992, and 2.742, respectively. The data measured by HAWC for Geminga is plotted as well (cyan). Top right: $\gamma$-ray flux found for different energy losses described by a single power law with $\alpha$ = 1.9 and different normalizations $b_0$ from $1.0 \times 10^{-17}$ to and $2.0 \times 10^{-16}$ GeV$^{-1.9}$. Bottom left: $\gamma$-ray flux with different values of $E_c$ from 100 GeV to 10000 TeV. Bottom right: $\gamma$-ray flux obtained by varying $\tau_0$ from 1 kyr to 1000 kyr.} 
        \label{fig:varying_parameters}
    \end{figure*}

Figure \ref{fig:varying_parameters} shows the $\gamma$-ray flux calculated by varying some of the physical parameters. The benchmark model adopts $\delta=1/3$, $D_0 = 1.0 \times 10^{26} \mathrm{cm}^2~\mathrm{s}^{-1}$, $\theta = 10$ degrees, $\tau_0 = 12$ kyr, $E_c = 1000$ TeV, $\gamma_e = 2.0$, $\eta=0.10$, and energy loss described in Vernetto 2016 \cite{Vernetto:2016alq}.
To be consistent with the physical assumptions used to derive HAWC2017 data, a $\theta$ of $10$ degrees is assumed\footnote{The data measured by HAWC for the energy spectrum of $\gamma$-ray depends on morphological assumptions since different integrating regions are considered for different morphology. Since we are using the diffusion model, we choose data generated assuming the diffusion morphology.}.
The parameters that we vary in this part are $\gamma_e$, $\tau_0$, $E_c$ and energy losses.
We choose to change these parameters because they are the one that affect the most the flux shape.
Instead, we do not make any variation of $D_0$ because we have tested that the $\gamma$-ray energy spectrum is independent of $D_0$.

In the top left panel of Figure \ref{fig:varying_parameters}, we vary $\gamma_e$ from 1.8 to 2.3 and find the best-fit $\eta$ for each $\gamma_e$ with HAWC2017 data. 
The best-fit efficiency values vary from 0.08 for $\gamma_e=1.8$ up to 2.74 found with a much softer spectrum of 2.3.
%0.118, 0.198, 0.406, 0.992, and 2.742, respectively. 
We see that spectrum with larger $\gamma_e$ have a steeper flux of $\gamma$ rays since more low-energy $e^{\pm}$ are injected, causing less high-energy photons to be produced through ICS. For the same reason, higher efficiencies would be required for larger $\gamma_e$. 
%While gravitational wave and protons account for less than $10\%$ of the spin-down energy of pulsar, the maximum possible efficiency for $e^{\pm}$ should be at least a few percent less than 1 \cite{Recchia_2021, Abbott_2008, Bucciantini_2010}.
We verified that $\gamma_e$ greater or equal to 2.1 are unrealistic since they require efficiencies greater than 1.
Although different $\gamma_e$ produce equally good fit to HAWC 2017 data, the flux below 1 TeV is very different for different $\gamma_e$. 
This suggests that the precision of the data and energy coverage in HAWC 2017 is insufficient to constrain $\gamma_e$ effectively. 

While the dependency of $\gamma$-ray flux on spectral index is well-known, the effect of other physical parameters including the energy losses, $\tau_0$ and cutoff energy $E_c$, on energy spectrum has rarely been discussed explicitly. In the top right of Figure \ref{fig:varying_parameters}, we adopt energy losses described by a single power law with slope of 1.9 and we vary the value of $b_0$. A very large cutoff energy of $10^5$ TeV is used. The panel shows that more significant energy losses of $e^{\pm}$, represented by a larger $b_0$, cause less photons to be produced, especially at high energies. The flux of low-energy photons is affected as well even if at a less extent with respect to high-energy $\gamma$ rays. In fact the flux above 200 GeV differs by one to two orders of magnitude for a 10 times different $b_0$.  
The $\gamma$-ray flux shows the existence of a maximum $\gamma$-ray energy, where the flux falls almost vertically. We see that the maximum $\gamma$-ray energy is smaller with larger values of $b_0$. This is explained by the fact that the larger is $b_0$, the larger are the energy losses and the quicker very-high-energy electrons lose energy. Thus, the determination of the Galactic average energy loss is crucial to the success of prediction of parameters of the model.

In bottom left panel of Figure \ref{fig:varying_parameters}, we vary $E_c$ from 100 GeV to 10000 TeV. The maximum energy of $\gamma$ rays is greatly affected by the value of $E_c$. For values of $E_c<10$ TeV, the $\gamma$-ray spectrum falls significantly below the HAWC $\gamma$-ray energies. For $E_c > 1000$ TeV, the $\gamma$-ray flux falls at around 500 TeV. The flux of low-energy photons is almost the same for $E_c$ greater than 10 TeV, since the number of low-energy $e^{\pm}$ is insignificantly affected by increase in $E_c$.
Given the above mentioned effect of $E_c$ on the $\gamma$-ray flux, values of $E_c$ larger than a few tens of TeV are required to fit the HAWC data.

In the bottom right panel of Figure \ref{fig:varying_parameters}, we vary $\tau_0$ from 1 kyr to 1000 kyr. Larger $\tau_0$ means that more $e^{\pm}$ are injected at earlier times in the lifetime of the pulsar and would have experienced more significant diffusion and energy losses. Thus, as shown in the panel, spectrum with larger $\tau_0$ has more low energy photons below 1 TeV. For example, the photon flux below 100 GeV for $\tau = $1 kyr is more than one order of magnitude higher than for $\tau_0 =$ 30 kyr. On the other hand, the number of high-energy photons is barely affected by variation in $\tau_0$. Spectrum with $\tau_0 =$ 1000 kyr is physically improbable as the age of pulsars including Geminga measured is smaller than $\tau_0$. 

In summary, we can conclude that: 1) the spectral index of $e^+$ injection spectrum affects the slope of the $\gamma$-ray flux as well as the efficiency value needed to fit HAWC2017 data; 2) more significant energy losses reduce the total number of photons emitted, especially at high energies, so determination of energy losses needs to be done before evaluation of $\gamma$-ray data; 3) the cutoff energy affects the flux of high-energy photons, and cutoff above a few hundred TeV are needed to produce $\gamma$ rays at the energies detected by HAWC; 4) a smaller characteristic spin-down timescale ($\tau_0$) of a pulsar increases the number of low-energy photons emitted, and smaller values can be obviously seen.

\subsection{Fit to HAWC2017 data for Geminga}
\label{sec:sb2017}

\begin{figure}
    \centering
    \includegraphics[width=0.45\textwidth]{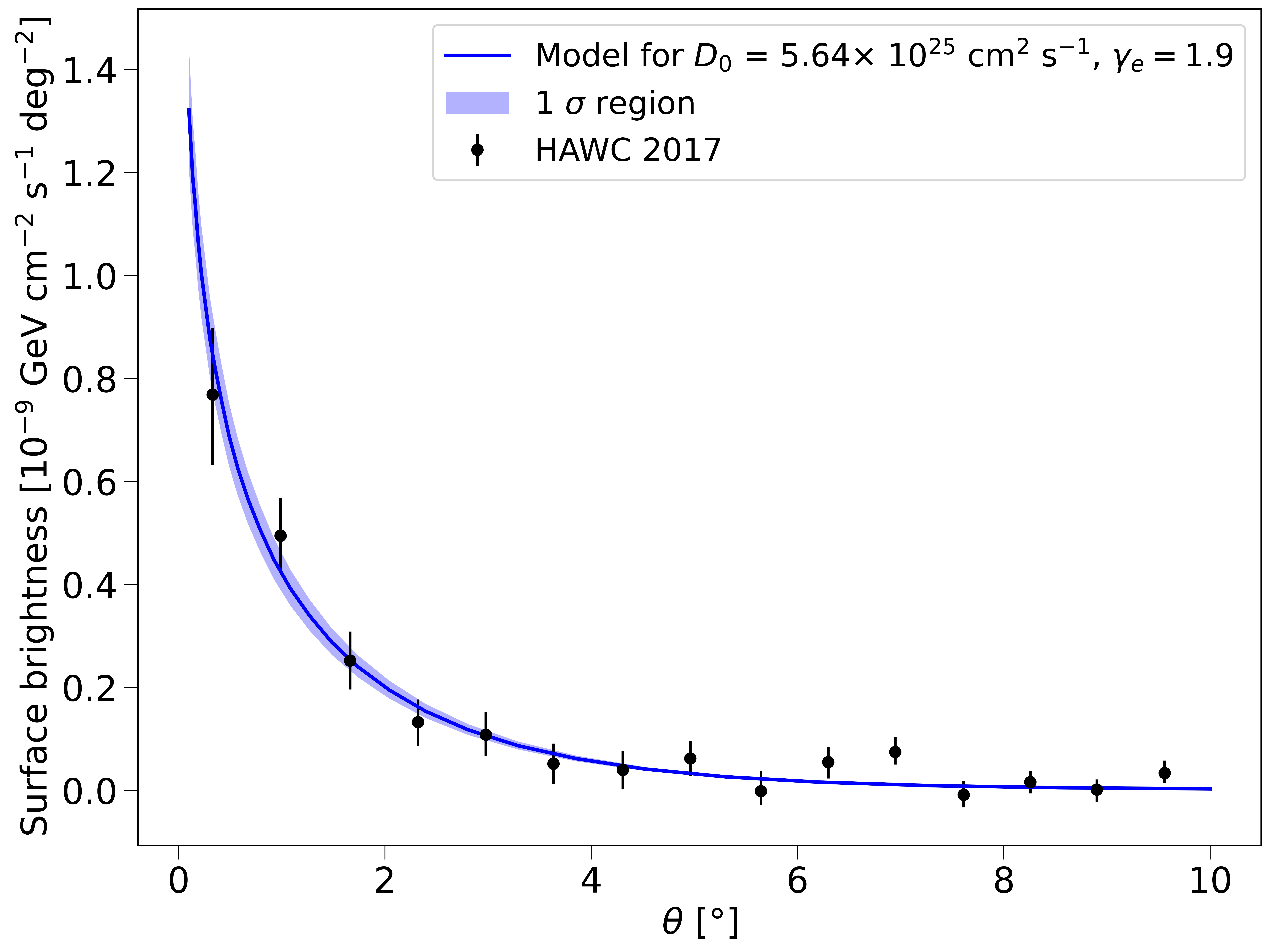}
    \caption{Best-fit surface brightness of Geminga fitted to HAWC2017 data, with $\gamma_e$ = 1.9, $\tau_0 = 12$ kyr, $E_c$ = 1000 TeV, $5<E_{\gamma}<50$ TeV, $\delta=1/3$, $D_0 = 5.64 \times 10^{25}~\mathrm{cm}^2 \mathrm{s}^{-1}$. 1$\sigma$ region due to uncertainty in efficiency is plotted. The HAWC2017 data (black) is plotted.}
    \label{Diff_2017}
\end{figure}

The model described in section \ref{Model} is used to fit the surface brightness of Geminga measured in HAWC2017 for $\gamma$ rays with energy $E_{\gamma}$ between 5 to 50 TeV.
In particular we vary in the model the efficiency, the normalization of the diffusion coefficient $D_0$. We also test how the results change using different values of $\gamma_e$. 
%Only the first eight data points are considered. 
We do not apply PSF correction to our theoretical predictions since it would affect marginally only the first data point. We again assume the benchmark parameters that are $\tau_0 = 12$ kyr, $E_{c} = 1000$ TeV. Figure \ref{Diff_2017} shows the best-fit surface brightness of Geminga when spectral index is assumed to be $\gamma_e =$ 1.9. In this case, the best-fit value of $D_0$ is $5.64^{+2.00}_{-1.54} \times 10^{25}~\mathrm{cm}^2 \mathrm{s}^{-1}$, with an efficiency of $\eta = 0.10 \pm 0.01$. 
The fit gives a chi-square of $\chi^2 = 2.78$ demonstrating the compatibility of the diffusion model with the observed surface brightness data. The HAWC2017 paper found diffusion factor at 100 TeV $D_{100}$ for Geminga to be $3.2^{+1.4}_{-1.0} \times 10^{27}~\mathrm{cm}^2 \mathrm{s}^{-1}$. This corresponds to $D_0 = 6.9^{+3.0}_{-2.2} \times 10^{25}~\mathrm{cm}^2 \mathrm{s}^{-1}$, since HAWC2017 adopts $\delta=1/3$. Ref.~\cite{Mattia_2019} found similar value of diffusion coefficient of $4.3^{+1.5}_{-1.2} \times 10^{25}~\mathrm{cm}^2 \mathrm{s}^{-1}$ for Geminga. We also demonstrate that our result is compatible with previous studies as the best-fit value of $D_0$ found lies in uncertainty band of both papers.

\begin{table}[ht]
\caption{Results of the fit to surface brightness of HAWC2017 data. The table shows the best-fit values of $D_0$ and $\eta$ obtained for different assumptions for $\gamma_e$. We also show the $\chi^2$ value obtained for each case.}
\centering
\begin{tabular}{c c c c c} 
\hline\hline 
$\gamma_e$ & 1.8 & 1.9 & 2.0 & 2.3 \\ [0.5ex]
\hline 
$D_0$ [$10^{25}~ \mathrm{cm}^2~\mathrm{s}^{-1}$] & $5.64^{+1.99}_{-1.70}$ & $5.64^{+2.00}_{-1.54}$ & $5.94^{+2.22}_{-1.70}$ & $6.41^{+2.63}_{-1.83}$ \\ 
$\chi^2$ & 2.83 & 2.78 & 2.81 & 2.83 \\
$\eta$ & 0.075 & 0.101  & 0.166 & 2.11 \\ [1ex] % [1ex] adds vertical space
\hline
\end{tabular}
\label{tbl:2017}
\end{table}

Table \ref{tbl:2017} shows the results for the fit of surface brightness (HAWC2017) when $\gamma_e =$ 1.8, 1.9, 2.0 and 2.3, with benchmark values for all other parameters. We see that the best-fit diffusion coefficient values is not altered much by the change of $\gamma_e$ (within 1$\sigma$ region), with similar values of $\chi^2$.
The best-fit efficiency values vary from 0.075 for $\gamma_e=1.8$ up to 2.11 found for $\gamma_e = 2.3$. Although $\eta$ found for fit to the surface brightness is relatively smaller than $\eta$ found for fit to the energy spectrum, the discrepancy is rather small and they are overall compatible with each other. We find again that for $\gamma_e \geq 2.1$, unrealistic efficiency larger than 1 is required. This verifies that $\gamma_e$ must be smaller than about 2.1.
In order to obtain the precise value of $\eta$, we will conduct a fit flux. Although the efficiency derived from surface brightness is more model independent, as the data does not depend on morphological assumptions, we see that $\eta$ obtained from fit to the surface brightness and to the energy spectrum is compatible with each other.

\subsection{Fit to the flux data of Geminga}
\label{sec:es2021}

Ref. \cite{Escobedo_2021} provides new data for  the flux of Geminga measured by the HAWC collaboration, for $\gamma$-ray energies above 1 TeV within $\theta$ of 8 degrees around the source. These data, if combined with {\it Fermi}-LAT measurements between 10 GeV and 1 TeV, could constrain the $e^{\pm}$ emission mechanisms across 4 orders of magnitude. These data thus permit a more precise evaluation of the physical parameters described in section \ref{sec:varp}.
We conduct independent fit to the $\gamma$-ray energy spectrum for the following two different setups: \textit{setup 1}: a fixed value for $E_c$ at 100 TeV, and usage of different values of $\tau_0$; \textit{setup 2}: a fixed value for $\tau_0 = 30$ kyr and a variation of the values for $E_c$.
In both cases the parameters left free in the fit are $\gamma_e$ and $\eta$. In this way, we derive the best-fit values of $\gamma_e$ and $\eta$ as well as best-fit range of $\tau_0$ and $E_c$ at the same time. We adopt $D_0 = 1.0 \times 10^{26}~\mathrm{cm}^2~\mathrm{s}^{-1}$, $\delta=1/3$ and $\theta=30$ degrees \footnote{The Fermi-LAT measurement used a theta of 30 degrees to avoid leakage of photons at low energy. To be compatible with Fermi-LAT data, the data provided by HAWC collaboration has been adjusted to the same region.}. Upper limits are not included in the fits since the number of existing data points is good enough to constrain the spectrum without involving uncertainty in the upper limits. 

\begin{figure*}[hbt!]
    \includegraphics[width=0.45\textwidth]{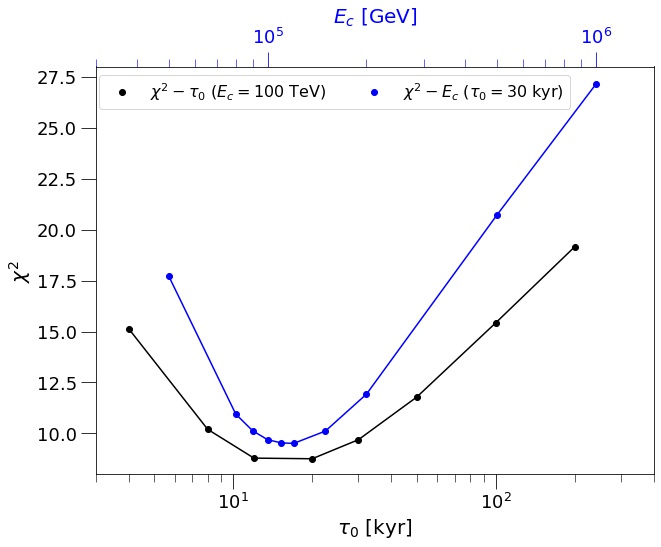}
    \includegraphics[width=0.49\textwidth]{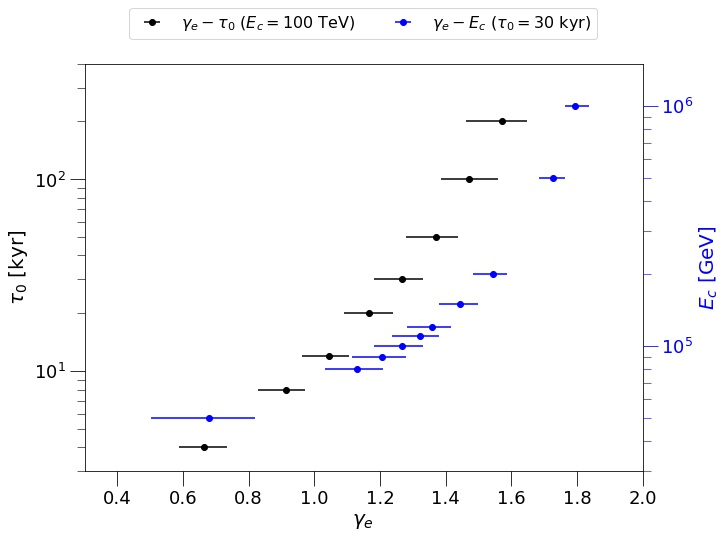}
    \caption{Left panel: Best-fit $\chi^2$ through fit to the $\gamma$-ray energy spectrum of Geminga for: setup 1: $\tau_0 = 30$ kyr and varying $E_c$ (blue, top x-axis) and setup 2: $E_c=100$ TeV with varying $\tau_0$ (black, bottom x-axis). Right panel: Best-fit $\gamma_e$ derived through fit to the energy spectrum of Geminga for setup 1 (black, left y-axis)  and setup 2 (blue, right y-axis). The efficiency increases with increasing $\gamma_e$ and is between $7.5-12.5\%$. We adopt $D_0 =1.0 \times 10^{26}~\mathrm{cm}^2~\mathrm{s}^{-1}$, $\delta = 1/3$ and $\theta = 30$ degrees for the fits.}
    \label{fig:fit-2021}
\end{figure*}

\begin{figure*}[hbt!]
    \includegraphics[width=0.49\textwidth]{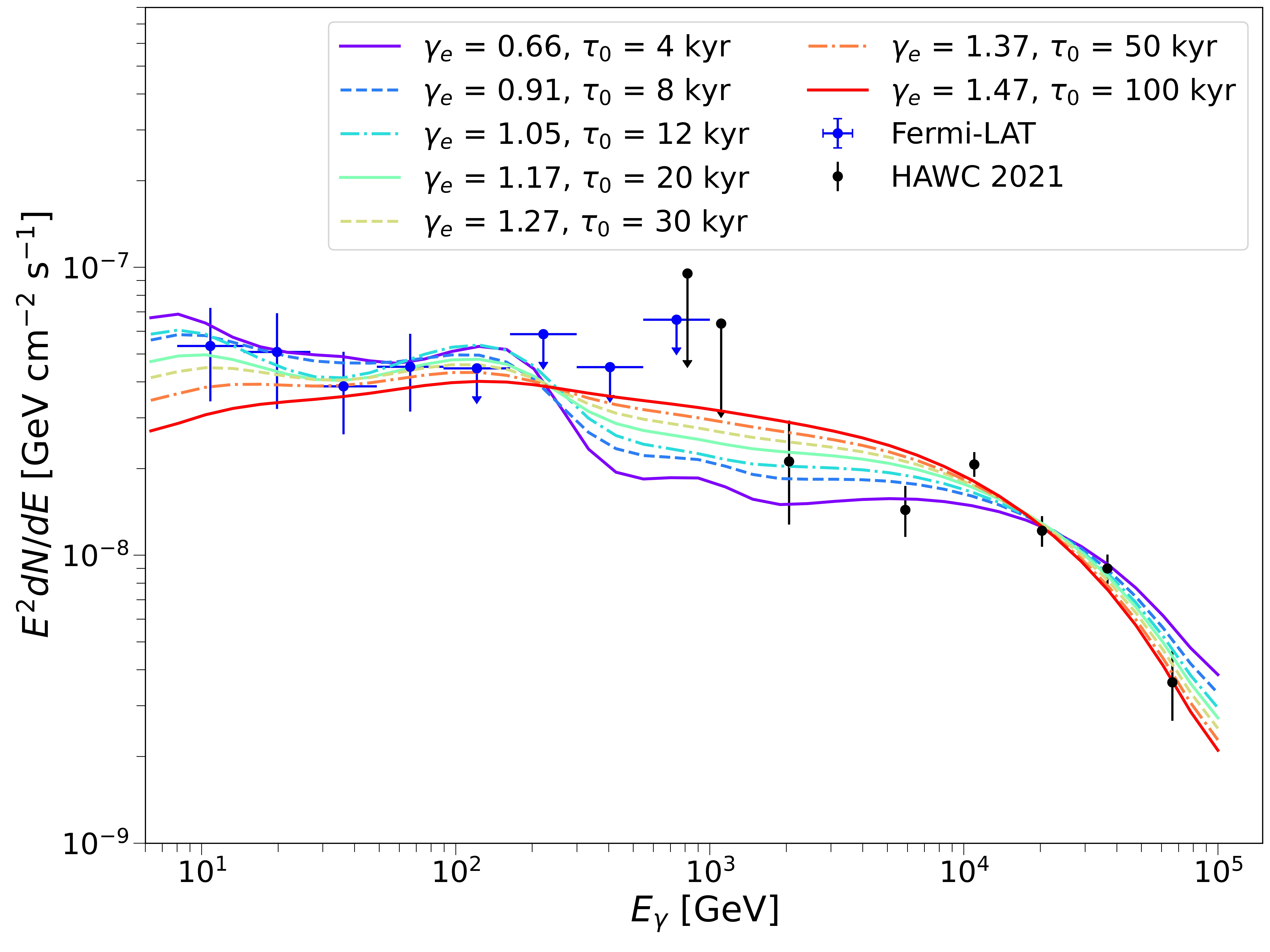}
    \includegraphics[width=0.49\textwidth]{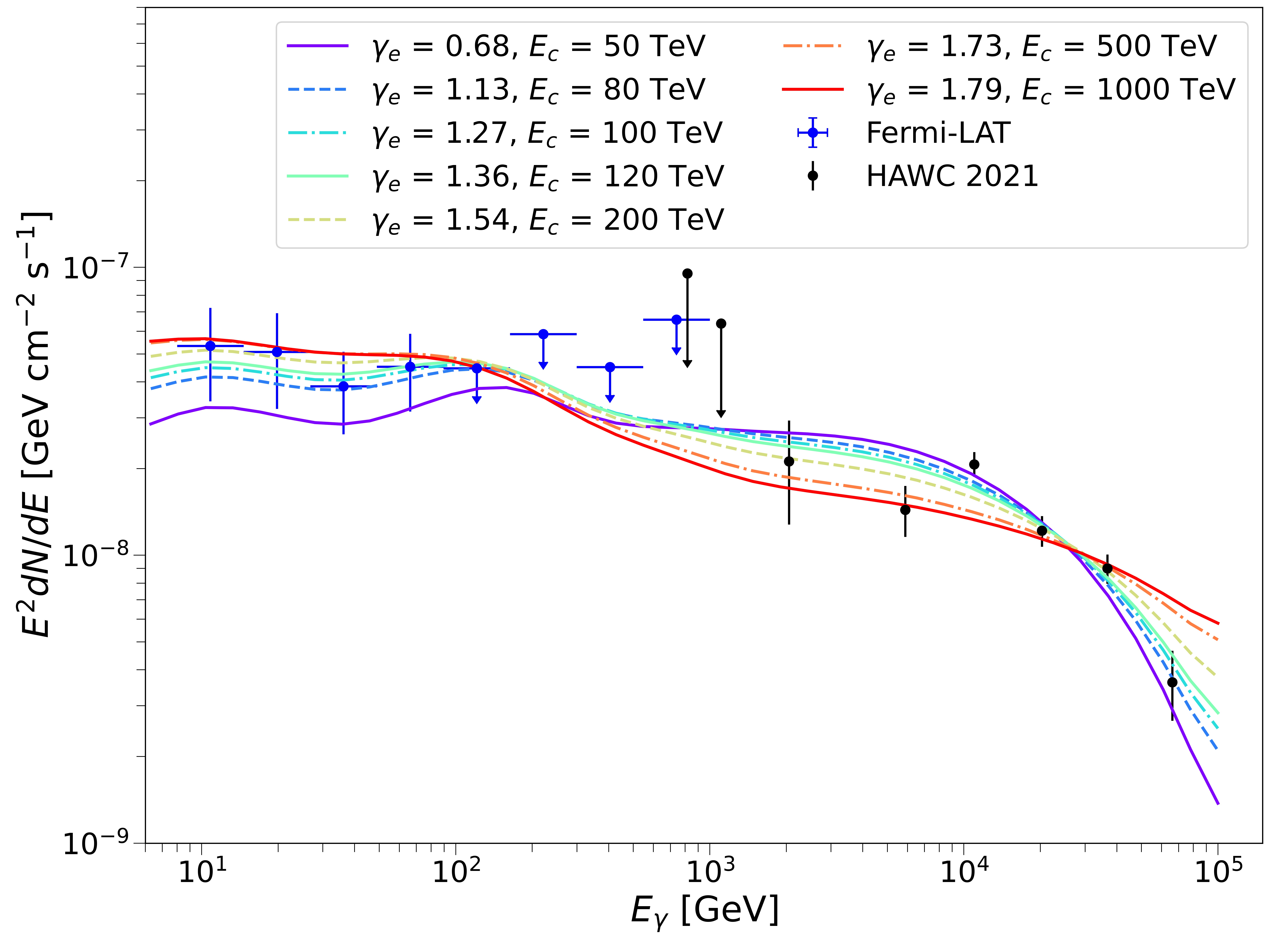}
    \caption{Left panel: Best-fit energy spectrum of Geminga for $E_c=100$ TeV and varying $\tau_0$ (setup 1), with best-fit $\gamma_e$ and $\eta$ derived from the fit. Right panel: Best-fit energy spectrum of Geminga for $\tau_0 = 30$ kyr and varying $E_c$ (setup 2), with best-fit $\gamma_e$ and $\eta$ derived from the fit.  We adopt $D_0 =1.0 \times 10^{26}~\mathrm{cm}^2~\mathrm{s}^{-1}$, $\delta = 1/3$ and $\theta = 30$ degrees for the fits. Data of Fermi-LAT (blue) and HAWC 2021 (black) are plotted.}
    \label{fig:Es2021}
\end{figure*}

Left panel of figure \ref{fig:fit-2021} shows the variation of best-fit $\chi^2$ for setup 1 (black line) and setup 2 (blue line).
With current precision of data, we find that the best range of $\tau_0$ and $E_c$ are $10-30$ kyr and $90-120$ TeV, respectively. The result with $\tau_0$ agrees with the usually assumed value of $\tau_0 = $ 12 kyr, which is derived in Ref.~\cite{Aharonian_1995}. The value of $E_c$ found is 10 times smaller than the assumed 1000 TeV in Ref.~\cite{Mattia_2019}. This is because, with more precise data, we see that the $\gamma$-ray flux falls significantly at $E_{\gamma} = 100$ TeV, matching a cutoff around 100 TeV. 
We find that $\chi^2$ increases rapidly once $E_c$ is outside the range of $90-120$ TeV, meaning that the cutoff energy is strictly constrained within this range. On the other hand, while $\tau_0$ of 12-20 kyr produces smallest value of $\chi^2$, the fit is less dependent on $\tau_0$ and a larger range of values is possible.

In the right panel of figure \ref{fig:fit-2021} we show the variation of best-fit $\gamma_e$ for setup 1 (black line) and setup 2 (blue line). We find that the best-fit $\gamma_e$ increases with increasing $\tau_0$ or $E_c$. This is expected as harder $\gamma_e$ values are needed to compensate the larger number of low-energy photons present in the flux when a smaller $\tau_0$ are assumed. The same thing applies when we reduce $E_c$, i.e. we have less high-energy photons. The efficiency increases with increasing $\gamma_e$ and it is between $7.5-12.5\%$.
When $\tau_0 = 12$ kyr and $E_c = 100$ TeV, the best-fit $\gamma_e$ is $1.05^{+0.06}_{-0.08}$ ($\chi^2 = 8.79$), with efficiency $0.091\pm0.024$. When $\tau_0 = 30$ kyr and $E_c = 100$ TeV, the best-fit $\gamma_e$ is $1.27^{+0.06}_{-0.09}$ ($\chi^2 = 9.69$), with efficiency $0.108\pm0.024$.

Figure \ref{fig:Es2021} shows the best-fit $\gamma$-ray flux for setup 1 (left panel) and setup 2 (right panel). Even if upper limits are not considered in the fit, we find that most of them are compatible with the model. However, to satisfy all the upper limits, especially the one around 100 GeV, $\tau_0$ of 30 kyr or beyond is required. Thus, although $\tau_0$ around 12-20 kyr produces the smallest value of $\chi^2$, the values of $\tau_0$ remain uncertain. 

HAWC2017, Ref.~\cite{Mattia_2019} and section \ref{sec:varp} shows that the range of $\gamma_e$ that fit HAWC2017 data well are between 2.0 and 2.3. Similarly, Ref. \cite{Fang_2018} uses $\gamma_e =2.2$ based on HAWC2017 data. In comparison, the best-fit spectral index we found through the fit to the new flux data is significantly smaller.
Considering that the data measured by HAWC in 2021 is larger by a factor of 2 than that in 2017, a smaller spectral index is reasonable. In addition, a smaller $\gamma_e$ is needed to compensate the choice of a smaller cutoff energy value of 100 TeV. %Our result is partially compatible with previous studies as well.
The fact that the best-fit spectral index in this work is harder than the index below the break in the broken power-law (which equals 1.5 in Ref.\cite{Tang_2019} and 1.8 in Ref. \cite{Evoli_2021}) suggests that: 1) a broken power law with a spectral break at a few hundreds of GeV might not be compatible with $\gamma$-ray emissions from Geminga; 2) the previously proposed broken power law might not be true for all pulsars. However, it should be noted that the current $\gamma$-ray data only constrains the TeV energies of the $e^+$ injection spectrum and more studies are needed to confirm this result.
%A recent paper has found a hard spectral index of [VALUE] through [DESCRIPTION] [CITATION]. 
The efficiency we found is lower than the result of HAWC2017 as well, considering smaller value of $\gamma_e$. While previous studies claim that gravitational waves account for less than $\approx 6\%$ of spin-down luminosity and protons account for at most a few percent \cite{Recchia_2021, Abbott_2008, Bucciantini_2010}, the fact that production of $e^{\pm}$ account for only $10\%$ of the spin-down luminosity suggests that either gravitational wave and ions are responsible for higher proportion of spin-down luminosity or there exists other unknown mechanism consuming significant portion of the energy.

\subsection{Fit to the surface brightness of Geminga}
\label{sec:sb2021}

The HAWC collaboration has measured new data of surface brightness of Geminga in 2020 (conference presentation) with higher precision in three energy bins, with $\gamma$-ray energy above 1 TeV, 5.6-17 TeV and 17-56 TeV (labelled hereafter HAWC 2020) \cite{Zhou_2020}. We adopt 1-50 TeV for the reconstructed energy bin above 1 TeV, as the contribution of photons with higher energy is relatively insignificant. For each energy bin, we fit the data for $D_0$ and $\eta$. Best-fit setup of physical parameters found in section \ref{sec:es2021} are used: $\tau_0 = 12$ kyr, $E_c = 100$ TeV, with corresponding $\gamma_e = 1.05$. 
%All data points measured by HAWC are included in the fit. 
The units of surface brightness data are given in counts per steradians and not in flux per angle. Therefore, in order to have an estimate of the efficiency we should apply convolve the model with the acceptance of the experiment. Since we are not interested in a precise estimation of $\eta$ through the surface brightness data, we consider them as random units. We remind that the $\eta$ parameter has been estimated before with the flux data.  
%it is simply rescaled from usual unit $\mathrm{GeV}~\mathrm{cm}^{-2}~\mathrm{s}^{-1}~\mathrm{deg}^{-1}$ by acceptance of measurement, which is independent of the angle. 
Hence, the fit of $D_0$ can be done with $\eta$ unconstrained. 
%The caveat may be that the efficiency obtained from energy spectrum is less model independent. However, discrepancy between fits of $\gamma$-ray energy spectrum and surface brightness shown in section \ref{sec:sb2017} is insignificant enough to be neglected.

The best-fit $D_0$ are found in the same way presented in section \ref{sec:sb2017}. For the same reason that the PSF correction only marginally affects the first data point, we do not apply it. The best-fit values of the diffusion coefficients are found to be:
\begin{itemize}
 \item $4.92^{+1.63}_{-1.11} \times 10^{25}~\mathrm{cm}^2 \mathrm{s}^{-1}$ ($\chi^2 = 14.09$), for 1 TeV $<E_{\gamma}<50$ TeV
 \item $6.24^{+3.26}_{-1.97} \times 10^{25}~\mathrm{cm}^2 \mathrm{s}^{-1}$ ($\chi^2 = 6.19$), for $5.6 < E_{\gamma} < 17$ TeV
 \item $1.05^{+0.49}_{-0.26} \times 10^{26}~\mathrm{cm}^2 \mathrm{s}^{-1}$ ($\chi^2 = 14.48$), for $17 < E_{\gamma} < 56$ TeV.
\end{itemize}
For different $\tau_0$ and $E_C$ within the best-fit range used, the results are very similar.

\begin{figure*}
        \centering
        \begin{subfigure}[b]{0.475\textwidth}
            \centering
            \includegraphics[width=\textwidth]{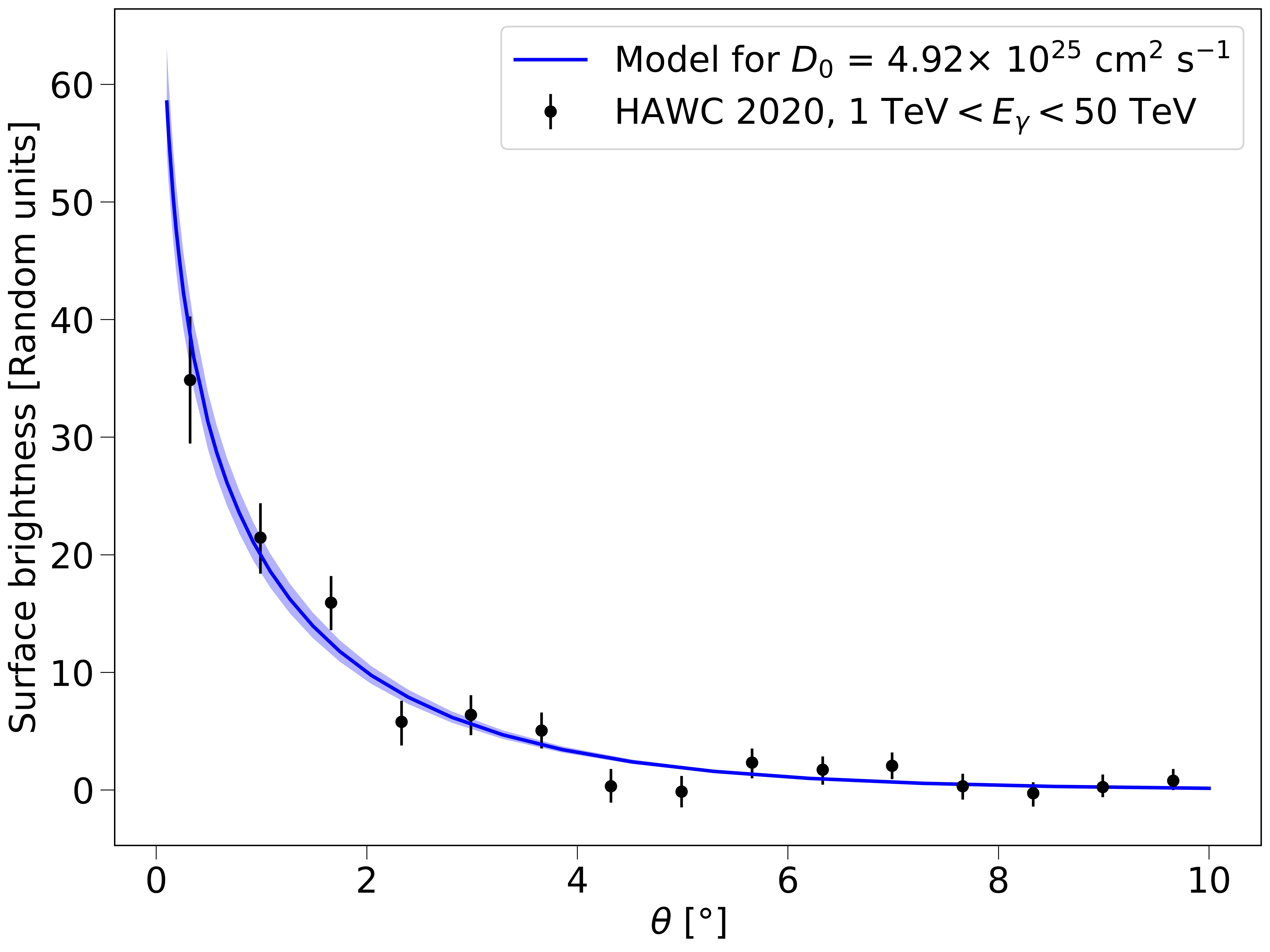}
        \end{subfigure}
        \hfill
        \begin{subfigure}[b]{0.475\textwidth}  
            \centering 
            \includegraphics[width=\textwidth]{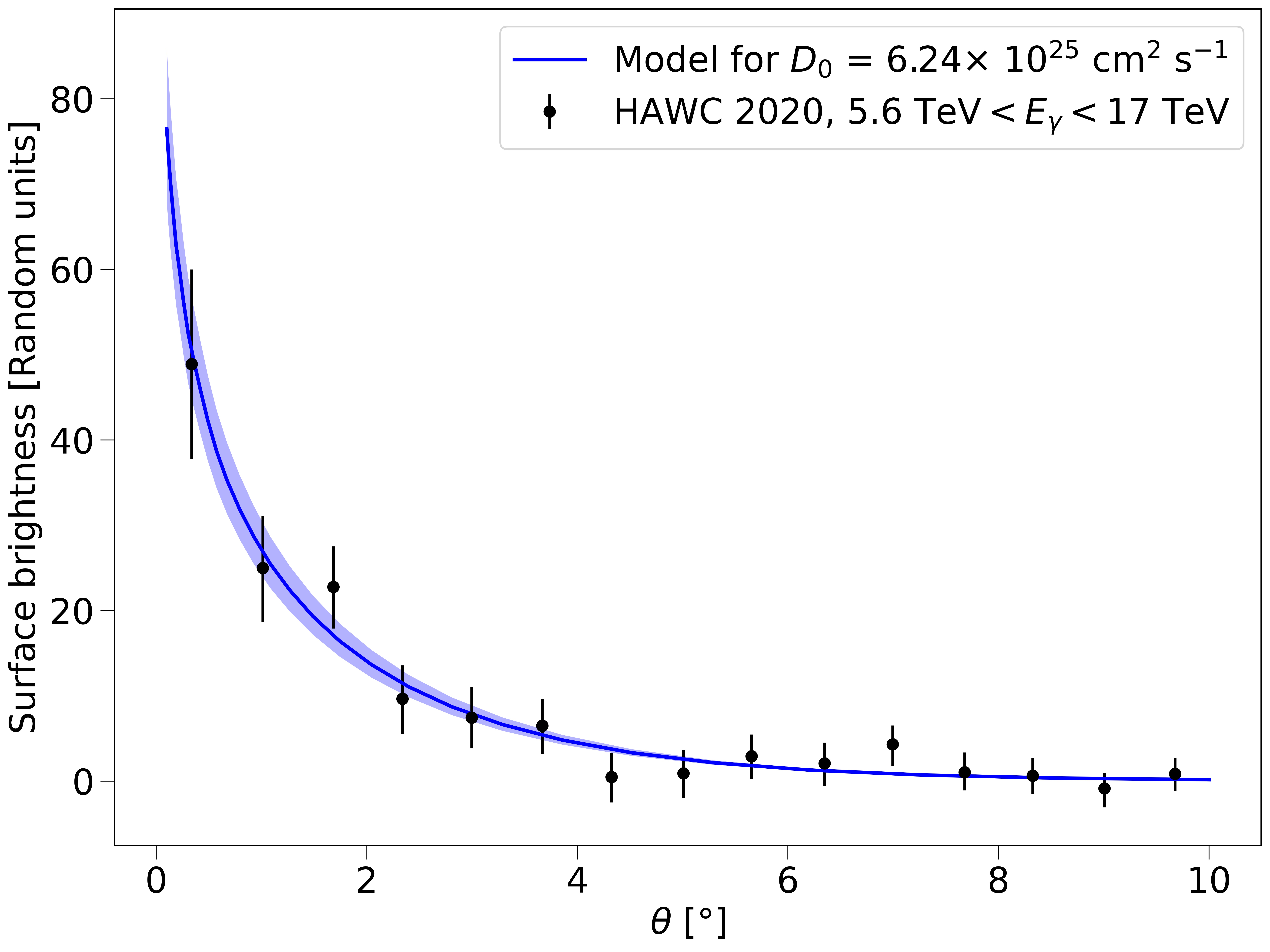}
        \end{subfigure}
        \vskip\baselineskip
        \begin{subfigure}[b]{0.475\textwidth}   
            \centering 
            \includegraphics[width=\textwidth]{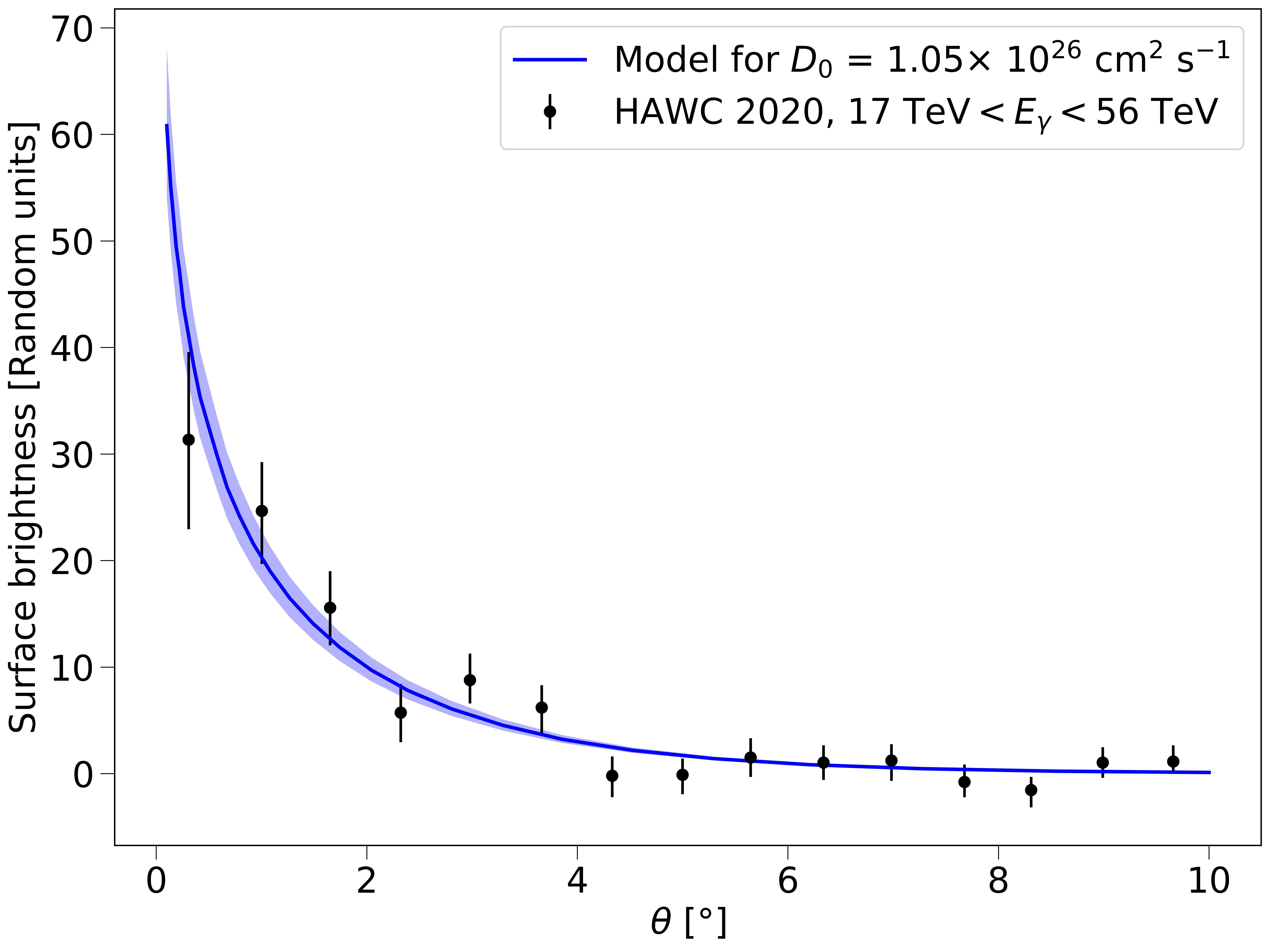}
        \end{subfigure}
        \hfill
        \begin{subfigure}[b]{0.475\textwidth}   
            \centering 
            \includegraphics[width=\textwidth]{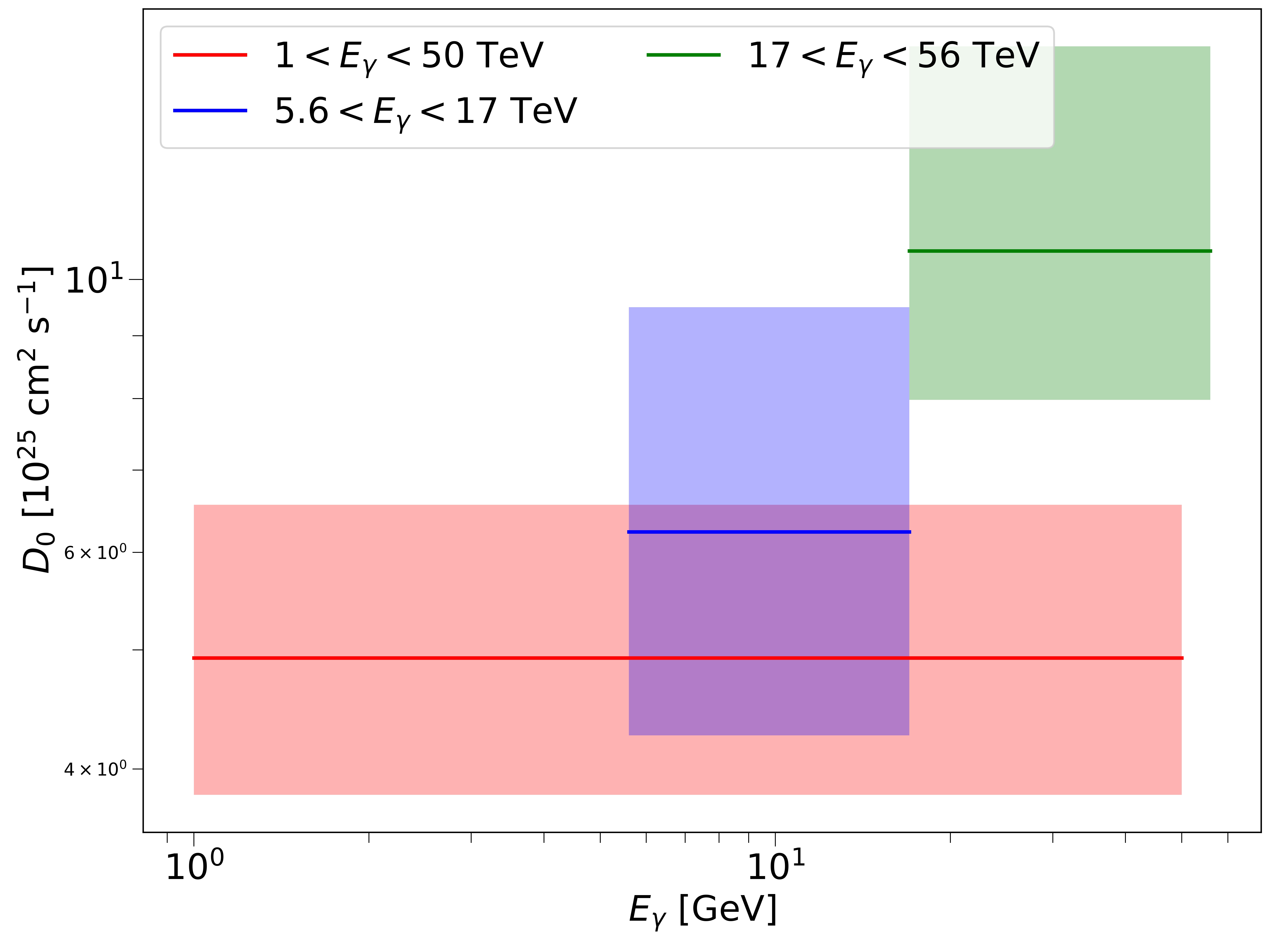}
        \end{subfigure}
        \caption[]
        {\small Results of the fit to the surface brightness of Geminga, with $\gamma_e = 1.05$, $\tau_0 = 12$ kyr, and $E_c = 100$ TeV. Top left: Best-fit surface brightness obtained for $1 < E_{\gamma} < 50$ TeV, with $D_0 = 4.92 \times 10^{25}~\mathrm{cm}^2 \mathrm{s}^{-1}$: Best-fit surface brightness of Geminga for $5.6 < E_{\gamma} < 17$ TeV, with $D_0 = 6.24 \times 10^{25}~\mathrm{cm}^2 \mathrm{s}^{-1}$. Bottom left: Best-fit surface brightness for $17 < E_{\gamma} < 56$, with $D_0 = 1.05 \times 10^{26}~\mathrm{cm}^2 \mathrm{s}^{-1}$. For these three panels, the HAWC2020 data is plotted, as well as $1\sigma$ region due to error in efficiency. Bottom right: The best-fit values of $D_0$ with 1$\sigma$ error band for different energy bins of $\gamma$-ray data. } 
        \label{fig:result_sb2021}
    \end{figure*}

The top left, top right and bottom left panels of Figure \ref{fig:result_sb2021} shows the best-fit surface brightness within 10 degree near Geminga for different energy bins. The values of $D_0$ found, especially for the first two energy bins, are consistent with data from HAWC2017. This shows that the diffusion model is compatible with the new data in different energy bins, with acceptable value of $\chi^2$ considering the number of data points.
It is confirmed that a diffusion coefficient that about two orders of magnitudes lower than the Galactic average value is present around Geminga. \par

The bottom right panel of Figure \ref{fig:result_sb2021} shows the values of $D_0$ with $1\sigma$ errors obtained for the different energy bins of $\gamma$-ray data. Despite the fact that the values of $D_0$ found are different for each energy bins, they are consistent with each other at $1\sigma$. The value of best-fit $D_0$ increases with the $\gamma$-ray energy. This may suggest that the value of $\delta$ is higher than $1/3$ that we used.
When we apply the most recent Galactic average value of $\delta$ of 0.51 \cite{AMS-02_new_parameters}, with $\tau_0 = 30$ kyr and $E_c = 100$ TeV, we find that the diffusion coefficients for different energy bins are more compatible with each other at $1\sigma$:
\begin{itemize}
    \item $7.96^{+2.11}_{-1.97} \times 10^{24}~\mathrm{cm}^2 \mathrm{s}^{-1}$ for $E_{\gamma} > 1$ TeV
    \item $9.96^{+5.17}_{-3.24} \times 10^{24}~\mathrm{cm}^2 \mathrm{s}^{-1}$ for 5.6 TeV $< E_{\gamma} <$ 17 TeV
    \item $1.57^{+0.56}_{-0.47} \times 10^{25}~\mathrm{cm}^2 \mathrm{s}^{-1}$ for 17 TeV $< E_{\gamma} <$ 56 TeV
\end{itemize}
For the $D_0$ of each energy bin to be strictly equal with each other, $\delta = 1.02 \pm 0.10$ would be required. 
This inconsistency may suggest that the inner diffusion zone not only has different diffusion coefficient but also different $\delta$. However, considering relative large errors of $D_0$, no firm conclusion can be made. This finding indicates that data with higher precision and more energy bins could be useful in verifying the diffusion model assumption and investigating the value of $\delta$ within the inner diffusion zone.

\subsection{Fit to flux data of J0621+3755}
\label{sec:LHAASOES}

\begin{figure}
    \centering
    \includegraphics[width=0.45\textwidth]{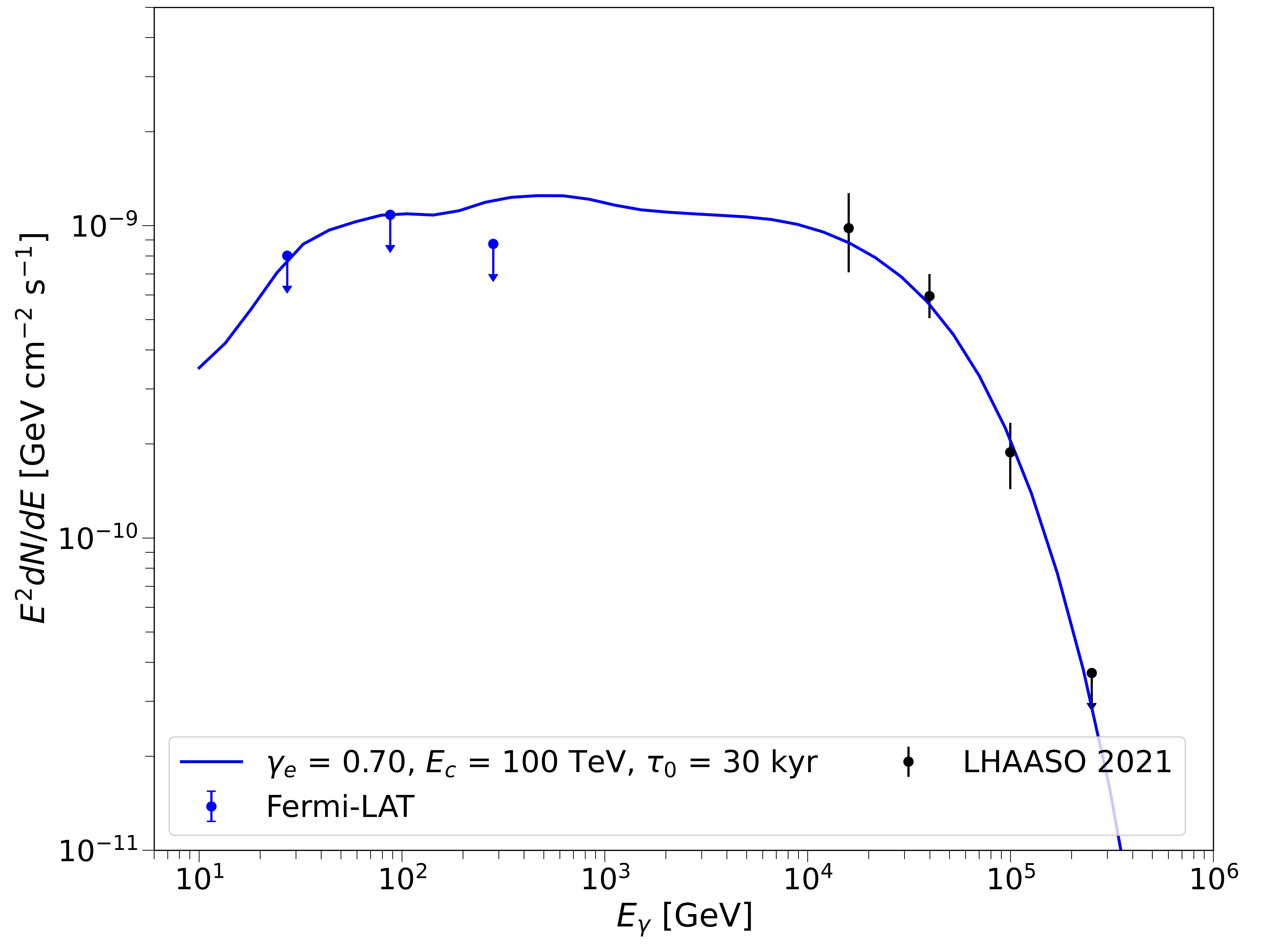}
    \caption{Best-fit flux for the source J0621+3755 for 10 GeV $< E_{\gamma} < 10^6$ GeV, with $\gamma_e = 0.70$, $\eta = 0.20$, $\tau_0 = 30$ kyr, $E_c = 100$ TeV, $D_0 = 1.0 \times 10^{26} ~\mathrm{cm}^2~\mathrm{s}^{-1}$, and $\theta = 12.5$ degrees. The $\chi^2$ of the best fit is 1.52.}
    \label{esLHAASO}
\end{figure}

The analysis of the data for the pulsar J0621+3755 is included in this paper to compare the diffusion model obtained with this source with respect to the ones obtained in the last section with Geminga. In particular, this is important to verify whether the properties of $\gamma$-ray halos found in sections \ref{sec:es2021} and \ref{sec:sb2021} are unique to the Geminga pulsar or could be a common feature present around all Galactic pulsars. Ref.~\cite{Aharonian_2021} presents data measured by LHAASO for the $\gamma$-ray emission around the pulsar J0621+3755 at energies above 25 TeV. 
We conduct a fit to the LHAASO and {\it Fermi}-LAT flux data for the source J0621+3755. We use the same procedure as described in section \ref{sec:es2021}. Upper limits are considered since there is a limited number of data points. Since the region of interest measured by LHAASO is a square of size $25^\circ$ by $25^\circ$ centered at the pulsar position, we adopt $\theta = 12.5$ degrees for the fit.

Figure \ref{esLHAASO} shows the $\gamma$-ray flux of the pulsar J0621+3755. We find that the best-fit model has $\gamma_e = 0.70$, $\eta= 0.20$, $E_c = 100$ TeV and $\tau_0 = 30$ kyr. The fit has a chi-square of $\chi^2$ of 1.52. Similarly to the results for Geminga, the cutoff energy is strictly constrained within a small range near $100$ TeV. 
On the other hand, when $\tau_0 = 12$ kyr is used, instead of 30 kyr, we find $\gamma_e = 1.01^{+0.41}_{-0.48}$ and $\eta= 0.29$ ($\chi^2 = 3.73$). Also for this pulsar the best-fit range for $\tau_0$ is between 10 and 30 kyrs.
Instead, contrary to the case with Geminga, smaller $\gamma_e$ are found for a larger $\tau_0$. 
This is due to the comparably low values of Fermi-LAT upper limits at low energy. When $\tau_0 = 12$ kyr, a larger $\gamma_e$ around 1.01 satisfies the LHAASO data points the best, while the Fermi-LAT upper limits cannot be satisfied with any $\gamma_e$ because the theoretical flux becomes larger than the upper limits. When $\tau_0 = 30$ kyr, a smaller $\gamma_e$ helps satisfying the Fermi-LAT upper limits as well as the LHAASO data points.

In general, we find consistency of $E_c$ and $\tau_0$ for results found for J0621+3755 and for Geminga. The cutoff energy $E_c$ is around 100 TeV. When $\tau_0 = 30$ kyr, the data points as well as the upper limits are satisfied well. For both pulsars, the predicted flux with $\tau_0 = 12$ kyr does not satisfy the upper limits but fit well the LHAASO data points. 
The value of $\gamma_e$ for two pulsars are consistent for $\tau_0 = 12$ kyr to be around $1$. However, when $\tau_0 = 30$ kyr, a smaller $\gamma_e = 0.70$ is found for J0621+3755, where the upper limits at low energy is relatively lower, while a much larger $\gamma_e = 1.27$ is found for Geminga. In addition, the best-fit efficiency of J0621+3755 is around $0.20$ ($\tau_0 = 30$ kyr) or $0.29$ ($\tau_0 = 12$ kyr), which is twice as much as efficiency of Geminga.
The differences of the model parameters written above are probably reasonable given the uncertainties of the data and because Galactic pulsars could have intrinsic differences in the emission mechanism.
%Although LHAASO 2021 has adopted Gaussian assumption for the morphological assumption for energy spectrum, causing the flux to be inaccurate, the general conclusions should be true. 

The cutoff energy around 100 TeV found for both pulsars reveals that the maximum energy of $e^{\pm}$ injected could reach the PeV energies. 
Similar spectral indexes for the pulsars may reveal that the injection mechanism of different pulsars is similar. Our result about the injection spectrum of $e^+$ may help constrain the parameters in models for the structure of PWNe and injection mechanism of $e^{\pm}$, for example, Ref. \cite{Bucciantini_2005, Olmi_2015, Del_Zanna_2017}. 
%Theoretically, the value of $\tau_0$ is given by \cite{Gao_2017}:
%\begin{equation}
%    \tau_0 = \frac{P_0}{(n-1)\dot{P}_0},
%\end{equation}
%where $P_0$ is the initial spin period, $n$ is the magnetic braking index and $\dot{P}_0$ is the initial period derivative. This means initial spin-down conditions of the pulsars could be evaluated using $\tau_0$. BUT WE DO NOT KNOW $P_0, \dot{P}_0$.

\subsection{Fit to the surface brightness of J0621+3755}

\begin{figure}
    \centering
    \includegraphics[width=0.45\textwidth]{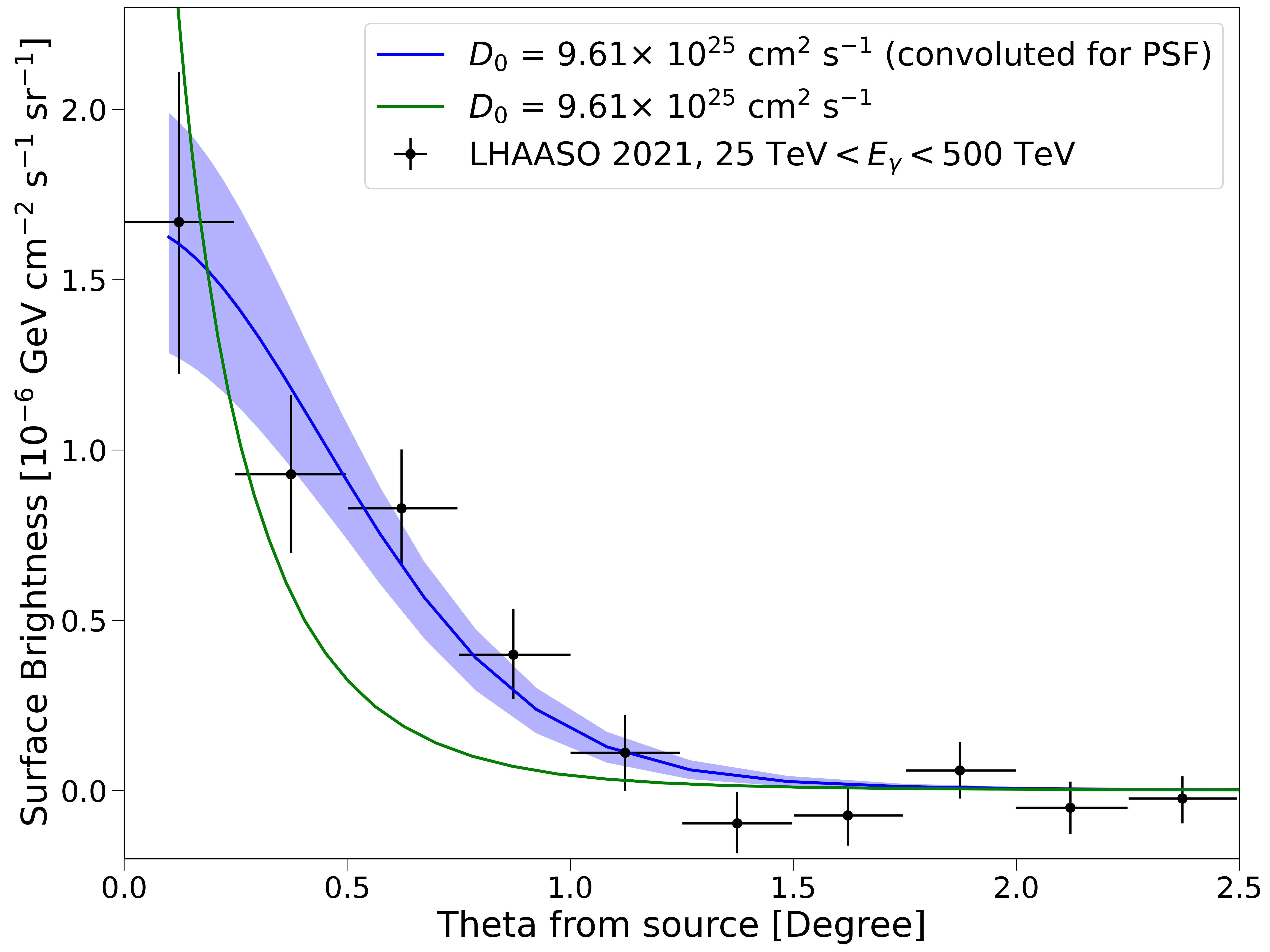}
    \caption{Best-fit surface brightness of the pulsar J0621+3755 corrected for the PSF (blue line) and uncorrected (green line). We also show in the same figure the data measured by LHAASO \cite{Aharonian_2021}. See the text for further information about the model parameters used. The 1$\sigma$ band is also plotted. The $\chi^2$ of the fit is 6.82.}
    \label{sbLHAASO}
\end{figure}

The surface brightness of the pulsar J0621+3755 is given for energies integrated above $E_{\rm{min}} = 25$ TeV.
%and $E_{\rm{max}} = 500$ TeV. 
The measurement is performed as a function of angle within a small region of $2.5$ degrees.
The point spread function (PSF) of LHAASO at these energies is around $0.4$ degree. Thus, unlike for Geminga, convolution of the surface brightness with the PSF needs to be done. We conduct the fit with best-fit parameters of the injection spectrum as found in section \ref{sec:LHAASOES}, i.e., $\gamma_e = 0.70$, $\tau_0 = 30$ kyr and $E_C = 100$ TeV. Figure \ref{sbLHAASO} shows the best-fit surface brightness corrected for PSF (blue) and uncorrected (green). The best-fit diffusion coefficient is $D_0 = 9.6^{+11.7}_{-6.2} \times 10^{25} ~\mathrm{cm}^2~\mathrm{s}^{-1}$. An efficiency of $\eta = 0.092\pm0.013$ is required by the fit and the chi-squared obtained with the best-fit is $\chi^2 = 6.82$. The uncertainties in the diffusion coefficient best-fit value is due to the convolution of the theoretical model with the PSF. In fact the convolution smoothens the model predictions make them similar to the PSF shape. Therefore, it reduces the effect of $D_0$ on the surface brightness profile. The efficiency found from the fit to the surface brightness is smaller by a factor of 2 than that found for $\gamma$-ray energy spectrum. However, considering that surface brightness is more model independent, such discrepancy is reasonable and $\eta$ determined through surface brightness is more reasonable. 

We find that the best-fit $D_0$ found near J0621+3755 in the energy bin 25 TeV$ < E_{\gamma} < 500$ TeV is consistent with the best-fit $D_0$ found around Geminga in the energy range between 17 TeV $ < E_{\gamma} < 56$ TeV. Since J0621+3755 has different position in space than Geminga, this result indicates that there might be a similar mechanism shared by pulsars causing inhibited diffusion near them. Large uncertainty of $D_0$ found near J0621+3755 means that it is compatible with that found for Geminga in all energy bins. In order to know better the diffusion coefficient, it is not only important to have a precise estimate of $D_0$ but also of the slope $\delta$.
Indeed, the propagation of cosmic rays is modelled mainly at the GeV-TeV energy range while $\gamma$-ray data are relevant only at energies well above 1 TeV. This implies that the diffusion coefficient slope is basically unconstrained from $\gamma$-ray observations. 
%In addition, the value of $D_0$ remains uncertain within $10^{25} - 10^{26} ~\mathrm{cm}^2~\mathrm{s}^{-1}$. This suggests more precise data of surface brightness of Geminga and J0621+3755 for different energy bins may be in need.
More observations of $\gamma$-ray data around several Galactic pulsars at GeV and TeV energies are needed to estimate precisely the Galactic diffusion coefficients and pulsar injection spectrum.

\subsection{Positron flux from Geminga}
\label{sec:posflux}

Section \ref{sec:es2021} and \ref{sec:sb2021} present different best-fit physical parameters for the injection of $e^{\pm}$ from Geminga derived from fits to $\gamma$-ray data. With the best-fit parameters previously found, we are able to compute the positron flux of Geminga using the two-zone diffusion model.
The efficiency value used for the positron flux is $\eta_{+} = \eta/2$, i.e.~one half of the one found from $\gamma$-ray data. Here we assume an equal number of positrons and electrons injected from pulsars. The time evolution of continuous injection is described by $\tau_0$; and the diffusion coefficient inside the $\gamma$-ray halo is equivalent to $D_0$. For the Galactic average $D_2$ of the outer diffusion zone, we use $1.17 \times 10^{28} \mathrm{cm}^2 \mathrm{s}^{-1}$ still with $\delta=1/3$ \cite{AMS-02_new_parameters}. 

\begin{figure*}[hbt!]
    \includegraphics[width=0.49\textwidth]{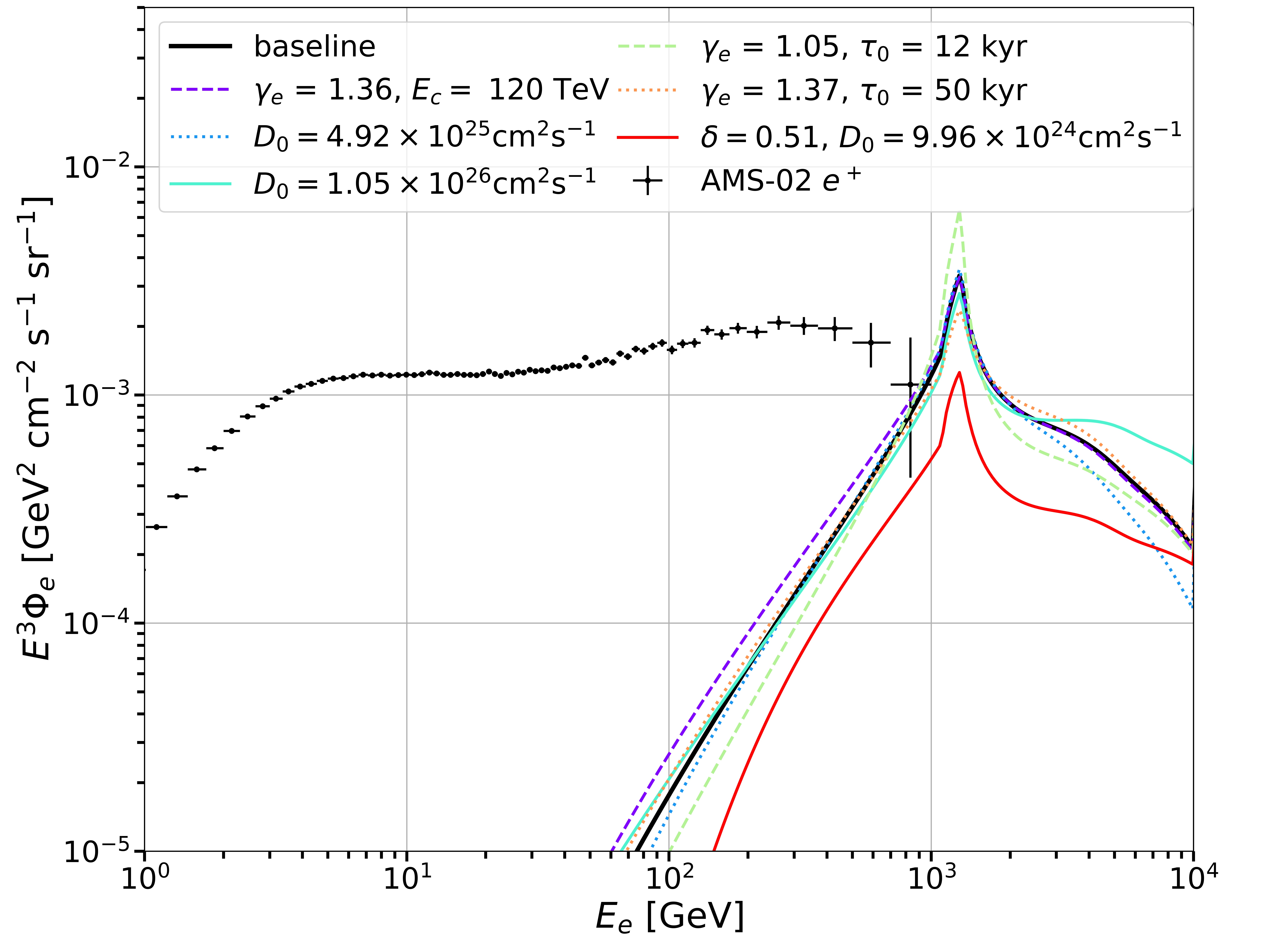}
    \includegraphics[width=0.49\textwidth]{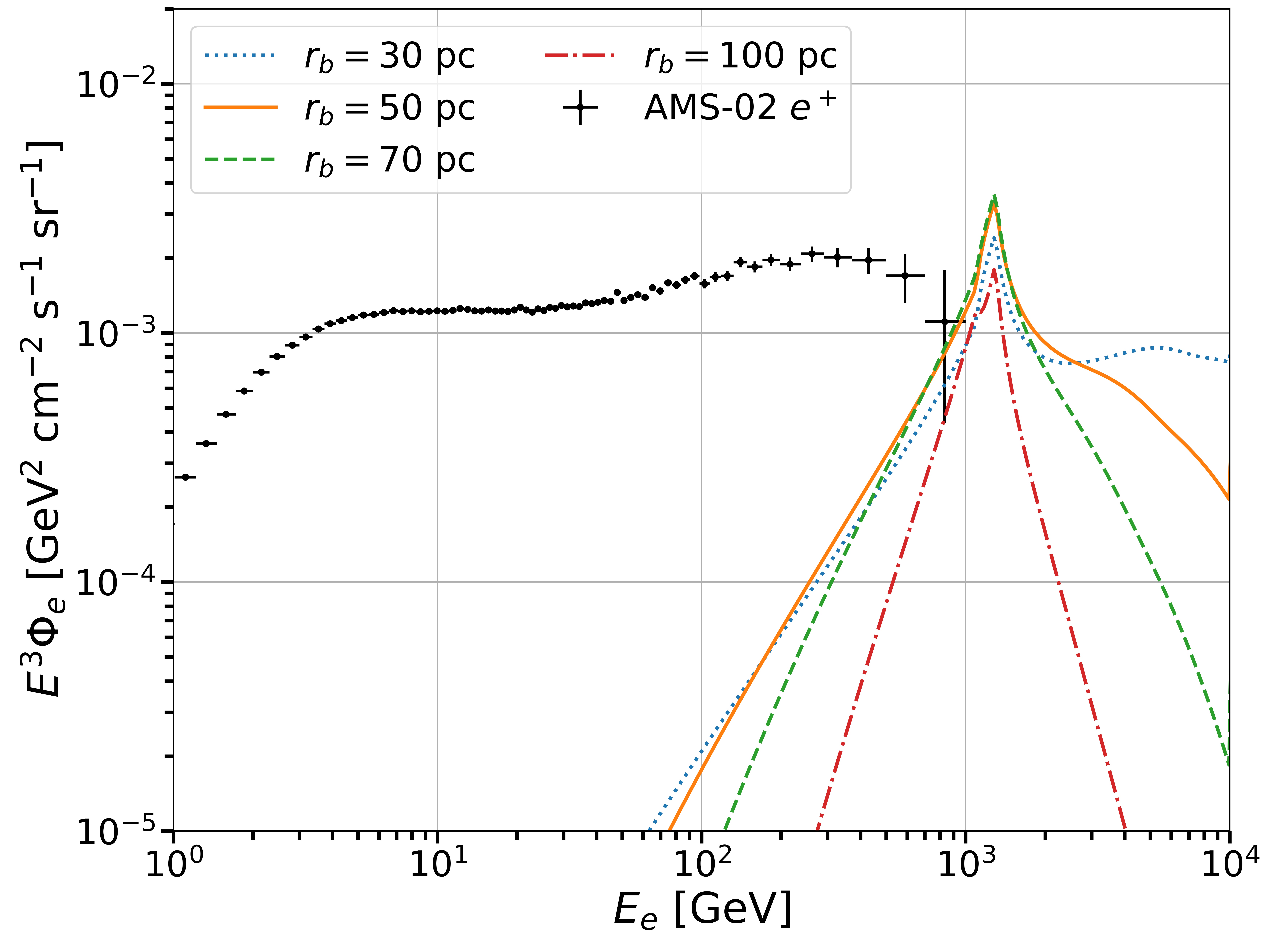}
    \caption{Contribution of Geminga to the positron excess based on two-zone diffusion model under different scenarios. Left panel: the positron flux of Geminga for $r_b = 50$ pc and different other parameters compared to the baseline models. The details of the baseline models can be found in the text. Additional lines represent the effect of different parameters on $e^+$ flux. Right panel: the positron flux of Geminga for different $r_b$, with the baseline values of parameters. The positron flux measured by AMS-02 is also shown in both panels of the figure. }
    \label{fig:posflux}
\end{figure*}

Figure \ref{fig:posflux} shows the $e^+$ flux of Geminga evaluated for different scenarios. 
The left panel shows the $e^+$ flux of Geminga with fixed $r_b = 50$ pc and different values of $\delta$, $\tau_0$, $E_c$, $D_0$ and corresponding best-fit values of $\gamma_e$ and $\eta_+$. The baseline values are $\tau_0 = 30$ kyr, $E_c = 100$ TeV, $\gamma_e = 1.27$ and $\eta_+ = 0.054$, and $D_0 = 6.24 \times 10^{25} \mathrm{cm}^2 \mathrm{s}^{-1}$ are motivated by the fit to $\gamma$-ray data.
Additional lines represent scenarios where different parameters are used. In particular we change the value of $\tau_0$ and $\gamma_e$ according to the best fit found in section \ref{sec:es2021} for the two parameters. For all scenarios, we find that Geminga makes significant contribution to the $e^+$ flux only for $e^+$ with energy greater than 100 GeV. The $e^+$ flux from Geminga increases at energy below 1 TeV, peaks at $E_e \approx$ 1 TeV and then falls for energy beyond a few TeV. 
The positron flux value at the peak is higher than the results of HAWC2017 by two orders of magnitude. The main reason for this difference is due to the probably wrong assumption that the diffusion coefficient from the source to the Earth is the inhibited one.
This assumption is probably wrong because if we consider that every pulsar in the Galaxy has an inhibited diffusion halo of the size of 200 pc, the entire Galaxy would have a diffusion coefficient of the order of $10^26~\mathrm{cm}^2 \mathrm{s}^{-1}$ that is not compatible with current CR observations.
Instead, our results are compatible with the ones of Refs.~\cite{Mattia_2019,Gulaugur_2019,Tang_2019}.

Since the values of $E_c$ and $\tau_0$ are uncertain, we show the $e^+$ flux for $\tau_0 = 12$ kyr, $\tau_0 = 50$ kyr or $E_c = 120$ TeV, with corresponding $\gamma_e$ and $\eta_+$ for each case. The $e^+$ flux with $\tau_0$ of 12 kyr is significantly more peaked at 1 TeV than the baseline case, while the flux with $\tau_0$ of 50 kyr is less peaked. The $e^+$ flux with $E_c =$ 120 TeV is almost the same as the flux of the baseline case, where the former has slightly larger flux for energy below 1 TeV.

While we use the $D_0$ near Geminga derived from the $\gamma$-ray data in the energy bin 5.6 TeV $<E_{\gamma}< 17$ TeV as baseline value of diffusion coefficient, we show as well the $e^+$ flux with $D_0$ derived in other energy bins with HAWC data. A smaller diffusion coefficient of $4.92 \times 10^{25} \mathrm{cm}^2 \mathrm{s}^{-1}$ causes the flux of high-energy $e^+$ to fall more quickly beyond 2 TeV, since $e^+$ are less likely to exit the inhibited diffusion zone. For the same reason, the $e^+$ flux with a larger $D_0 = 1.05 \times 10^{26} \mathrm{cm}^2 \mathrm{s}^{-1}$ is flatter above 2 TeV. Since the analysis of \citeauthor{AMS-02_new_parameters} shows that the Galactic average value of $\delta$ is around 0.51, we also compute the $e^+$ flux for $\delta=0.51$. For this case, we choose the corresponding diffusion coefficient of $D_0 = 9.96 \times 10^{24} \mathrm{cm}^2 \mathrm{s}^{-1}$ based on fit to the surface brightness near Geminga with $\delta=0.51$. The $e^+$ flux with $\delta=0.51$ is overall lower than that with $\delta=1/3$, since $e^+$ diffuse more while propagating through the Galaxy. In this case, the peak of the $e^+$ flux at 1 TeV is less significant, being very similar to the AMS-02 data at highest energy. 

Right panel of the Figure \ref{fig:posflux} shows the $e^+$ flux from Geminga for different radii of inner diffusion zone $r_b$. The value of $r_b$ cannot be determined through the current $\gamma$-ray data so we choose a few possible values including 30, 50, 70 and 100 pc. The overall shape of $e^+$ is similar for all cases, with a peak of the flux at around 1 TeV. We see that when a larger $r_b$ is used, the peak of $e^+$ flux is still at 1 TeV but it has a much narrower shape, since more $e^+$ are kept within the inhibited diffusion zone. For the case with $r_b = 70$ pc and 100 pc, the $e^+$ flux falls rapidly for energy greater than 1 TeV. When $r_b = 30$ pc is used, the $e^+$ flux continues to be flat for energy beyond 1 TeV. 

Recent papers undertaking simulations of Galactic pulsar populations find that the Galactic population of pulsars are able to explain the AMS-02 data well, by a few bright pulsars \cite{Cholis_2018, Orusa_2021, Evoli_2021, Cholis_2022}. On the other hand, Ref. \cite{Martin_2022} verifies that AMS-02 data may be explained well if development of diffusive halos is rare.
This agrees with our result, where we find the contribution of Geminga at highest energy of AMS-02 data is $30\%$ to nearly $100\%$. It also suggests that the contribution of Geminga to $e^+$ flux at higher energy (beyond 0.8 TeV) may be very significant. 
Consequently, we predict that the peak in $e^+$ flux from the pulsar Geminga at energy around 1 TeV would be clearly observed in future measurements of positron data. Based on our results, future measurement of $e^+$ flux at higher energy than AMS-02 may permit the precise determination of physical parameters for Geminga. The sharpness of the peak in $e^+$ flux may help constrain the value of $\tau_0$ for Geminga as well as the Galactic average value of $\delta$ more precisely. In addition, future $e^+$ flux data beyond 1 TeV may also help determine the precise value of $D_0$ and $r_b$ near Geminga, where the latter has always remained unconstrained.

The $e^+$ flux of J0621+3755 is not shown in this paper because the source is too far away from Earth, causing its contribution to be two-to-three orders of magnitudes smaller than the contribution of Geminga, depending on the physical assumptions on the injection parameters. The main contribution of J0621+3755 on $e^+$ flux is in the range of 1-10 TeV.

\section{Conclusions}
\label{sec:conclu}
The injection of $e^{\pm}$ from PWNe has been a promising candidate to explain the positron excess, which has been first detected by Pamela and then measured with great precision by AMS-02. The $\gamma$-ray emission near pulsars serve as indirect evidence for this explanation as these photons could be produced through inverse Compton scattering of $e^{\pm}$ against ISRF photons. The $\gamma$-ray data also help constrain key physical parameters for the injection, diffusion and propagation of $e^+$ from pulsars.

While the precision of previous data does not permit precise evaluation of the parameters of Geminga, in this paper, we analyze new $\gamma$-ray data for pulsar Geminga released by HAWC in 2020 and 2021 to estimate the parameters more precisely.
For the first time, we show explicitly the effect of varying different physical parameters, including $\gamma_e$, $E_c$, $\tau_0$ and energy losses, on the $\gamma$-ray energy spectrum.
%Our result is that the cutoff energy should be around 100 TeV, the spectral index around 1 and $\tau_0$ of a few tens of kyr
We conduct fit to the $\gamma$-ray energy spectrum of Geminga from 10 GeV to 100 TeV, with high-energy data from new HAWC measurement and low-energy data from Fermi-LAT measurement. We find the best-fit range of $\tau_0$ and $E_c$ for Geminga is 10-30 kyr and 90-120 TeV, respectively. The value of $E_c$ is constrained within a narrow range with an unprecedented precision. On the other hand, the value of $\tau_0$ remains uncertain, since $\tau_0$ greater than 30 kyr is required to satisfy all the upper limits while $\tau_0 = 12-20$ kyr produces smaller $\chi^2$ values. We observe that best-fit value of $\gamma_e$ increases with $\tau_0$ and $E_c$, while the best-fit efficiency remains in the range of $7.5\% - 12.5\%$. 
When $\tau_0 = 12$ kyr and $E_c = 100$ TeV, the best-fit $\gamma_e$ is $1.05^{+0.06}_{-0.08}$ ($\chi^2 = 8.79$), with efficiency $0.091\pm0.024$. 
When $\tau_0 = 30$ kyr and $E_c = 100$ TeV, the best-fit $\gamma_e$ is $1.27^{+0.06}_{-0.09}$ ($\chi^2 = 9.69$), with efficiency $0.108\pm0.024$. 

The value of $\tau_0$ found is consistent with previous estimate of 12 kyr \cite{Aharonian_1995}. The value of $E_c$ derived is smaller than the usual assumed 1000 TeV in Ref. \cite{HAWC2017, Mattia_2019}, as the flux falls significantly at around 70 TeV. The value of $\gamma_e$ found is also significantly smaller. This is because the high-energy photon flux data measured in 2021 is two times larger than the data of 2017, and a smaller $\gamma_e$ is required to compensate the choice of a smaller $E_c$. 
While protons and gravitational waves should account for less than $10\%$ of the total spin-down luminosity \cite{Recchia_2021, Abbott_2008, Bucciantini_2010}, our finding that the efficiency of injection of $e^{\pm}$ is only around $10\%$ reveals that either cosmic rays and gravitational waves account for greater proportion of energy or other unknown mechanism may be responsible for the consumption of the spin-down energy. Eventually other emission mechanisms should be responsible for the missing energy production.

We conduct also the fit to the HAWC data of the surface brightness of Geminga in three different energy bins, 1-50 TeV, 5.6-17 TeV and 17-56 TeV. 
The best-fit value of $D_0$ is found to be:
\begin{itemize}
    \item $4.92^{+1.63}_{-1.11} \times 10^{25}~\mathrm{cm}^2 \mathrm{s}^{-1}$ ($\chi^2 = 14.09$),
    \item $6.24^{+3.26}_{-1.97} \times 10^{25}~\mathrm{cm}^2 \mathrm{s}^{-1}$ ($\chi^2 = 6.19$),
    \item $1.05^{+0.49}_{-0.26} \times 10^{26}~\mathrm{cm}^2 \mathrm{s}^{-1}$ ($\chi^2 = 14.48$).
\end{itemize}
Inconsistency of $D_0$ for three energy bins may suggest a $\delta = 1.02 \pm 0.10$ is needed for the inner diffusive zone. However, considering the large errors in $D_0$, more precise measurement is in need to verify the result.

Analysis of PSR J0621+3755 is included in this paper to see whether result for Geminga is unique to Geminga or is a common feature present around Galactic pulsars. We conduct fit to the $\gamma$-ray flux for pulsar J0621+3755, with data measured by the LHAASO collaboration in 2021 and low-energy data from Fermi-LAT. We find that the values of $E_c$, $\tau_0$ and $\gamma_e$ for J0621+3755 are compatible with the result for Geminga, although low values of upper limits at low energy causes a smaller value of $\gamma_e$ of 0.70 to be found for $\tau_0 = 30$ kyr. This reveals that a similar injection mechanism may be shared by Galactic pulsars. 

We also conduct fit to the surface brightness for pulsar J0621+3755 in the range of 25 to 500 TeV, with data measured by LHAASO. A PSF correction is done during the fit since, unlike for Geminga, LHAASO measured surface brightness within 2.5 degrees from the source, with PSF of 0.40 degree. The best-fit value of $D_0$ is $1.08^{+1.1}_{-0.74} \times 10^{26} ~\mathrm{cm}^2~\mathrm{s}^{-1}$, which is consistent with the diffusion coefficient near Geminga found for the highest energy bin. This reveals that a common mechanism inhibiting diffusion may be shared by Galactic pulsars, so that the diffusion coefficients near pulsars are compatible, at least for $e^+$ with energy of tens of TeV. 

At last, we compute the $e^+$ flux from Geminga using the best-fit physical parameters derived in previous sections. We show that Geminga contribute to the $e^+$ flux for energy above 0.8 TeV significantly. We predict that the future measurement of $e^+$ flux would detect a sharp peak contributed by Geminga at energy around 1 TeV. We also show how different values of $\delta$, $D_0$, $\tau_0$, $E_c$ and $r_b$ would affect the $e^+$ flux. Based on our results, we believe that future measurement of $e^+$ at higher energy would also help constrain the parameters, especially the currently weakly constrained $\tau_0$ and $r_b$.

\section*{Acknowledgments}
The author would like to express his deepest thanks to Dr. Mattia Di Mauro. Without his assistance and supervision, this work would not have been possible.

\FloatBarrier
\bibliographystyle{apsrev4-1}
\bibliography{paper}

\end{document}